\documentclass[11pt]{article}
\pdfoutput=1

\usepackage{mymacros}
\title{Deep neural networks as lattice gauge theories}

\author{Ro Jefferson}
\author{and Shradha Ramakrishnan}
\affiliation{Institute for Theoretical Physics, and Department of Information and Computing Sciences,\\Utrecht University, Princetonplein 5, 3584 CC Utrecht, The Netherlands}

\abstract{We modify the NN/QFT duality \cite{Grosvenor:2021eol} to incorporate the layerwise permutation symmetry of the network, resulting in a $(0\!+\!1)$-dimensional lattice gauge theory, in which each layer of $N$ neurons acts as an $N$-component lattice site, and the weight matrices play the role of gauge fields living on the links. In this framework, we compute the tree-level neuron-neuron propagator which describes the evolution of layer variance in the network, and develop the Feynman diagram machinery to compute interactions in the perturbative expansion in $1/N$. In particular, we obtain a recursive expression for all corrections to the exact propagator at $O(1)$, representing statistical fluctuations in the ensemble of networks, including infinitely-many loop diagrams mediating the interactions from previous layers. We also present a preliminary analysis of neuron scattering amplitudes that contribute order-by-order in $1/N$, which provides a field-theoretic framework for studying higher-point correlations, and by extension information propagation, in deep networks. We remark on some interesting directions for future work at the intersection of neural networks and quantum field theory.
}

\begin{document}
\maketitle

\section{Introduction}

The past few years have witnessed a dizzying yet primarily empirically-driven advance of deep learning methods, popularized by the surprising success of large language models (LLMs). At the same time, there has been an explosion of work at the growing intersection of physics and machine learning, both with the goal of harnessing advances in AI for physics applications, and in the interest of applying ideas and techniques from physics to understand neural networks in the hopes of closing the gap between engineering practice and theory. Examples of the latter include applying path integrals to diffusion models \cite{Hirono:2024zyg}, energy-based methods for associative memory \cite{rooke2026stochasticthermodynamicsassociativememory} and transformers \cite{hoover2023energytransformer}, and diverse applications of dynamical mean field theory (DMFT) \cite{bordelon2026disordereddynamicshighdimensions,bordelon2023dynamicsfinitewidthkernel,hara2026dmftanalysishopfieldnetwork}. Well-known methods like the renormalization group have been applied to shed light on diffusion models \cite{Cotler:2023lem}, Bayesian learning \cite{Berman_2023}, and Gaussian process regression \cite{Howard_2025,coppola2026renormalizationgroupdeepneural}; see also \cite{greenspan2026worstcaseguaranteesscaleawareinterpretability} and references therein. More generally, statistical field theory has proven a powerful tool for studying large neural networks in various regimes of relevance to modern practice \cite{ringel2025applicationsstatisticalfieldtheory}. For a recent overview of other efforts at closing this gap, see \cite{simon2026scientifictheorydeeplearning}.

In this context, the intersection of neural networks and quantum field theory in particular has seen a number of works on the \emph{NN/QFT correspondence}. Broadly speaking, these are based on formal similarities between neural networks in the large-but-finite-width regime, and quantum field theories in Euclidean signature; in particular, when the number of neurons per layer $N$ is large, $1/N\ll 1$ serves as a parameter controlling a perturbative expansion that describes finite-width effects as one backs away from the idealized Gaussian process limit ($N\to\infty$). For pedagogical references on the application of such field-theoretic ideas to neural networks, see \cite{Roberts:2021fes,helias2019statistical}. More specifically, early works by Halverson and collaborators \cite{Halverson:2020trp,Halverson:2021aot,Demirtas:2023fir} developed a formal relation between the output of a neural network at large-width, treated as a functional, and a corresponding field theory, in which higher-point interactions in the action are controlled by $1/N$. This approach, dubbed \emph{neural network phenomenology} in \cite{Erbin:2021kqf}, has since been developed as a means of constructing familiar field theories -- such as $\phi^4$ theory, CFTs, and string theory -- by carefully designing a custom network so as to effect the desired properties, e.g., \cite{Halverson:2024axc,Frank:2025zuk,Frank:2026bui}. Collectively, this represents a new perspective on field theories through the lens of neural networks. 

Another approach within the NN/QFT umbrella, pioneered in \cite{Grosvenor:2021eol} (see also \cite{Segadlo_2022,zunigagalindo2026criticalorganizationdeepneural}) takes a complementary perspective: rather than engineer the output functional of a network to match the behaviour of known field theories, find a \emph{dual description} of a given network's hidden states in terms of a (potentially new) field theory. This approach can be traced back to early work by Sompolinsky et al. \cite{Sompo1988,Sompo1982}, and draws on well-established tools in statistical field theory, cf. the review \cite{helias2019statistical} mentioned above. The objective of \cite{Grosvenor:2021eol} in particular was to develop a dual description of DNNs (by which we mean fully-connected MLPs) and RNNs, and leveraged large-$N$ methods familiar from the study of $\mathrm{O}(N)$ vector models in high-energy theory to advance the diagrammatic analysis of perturbative effects via Feynman diagrams. These are relevant for, e.g., computing corrections to the critical point that determines optimal information propagation in phase space, which was hitherto only known in the idealized Gaussian process limit \cite{schoenholz2017deep}.

Here, we build on this work by incorporating the layerwise permutation symmetry of the DNN into the dual field-theoretic description, resulting in a lattice gauge theory with discrete symmetry group $S_N$. That fully-connected networks enjoy this symmetry is of course well-known in the machine learning literature (see for example \cite{brea2019weightspacesymmetrydeepnetworks,entezari2022rolepermutationinvariancelinear,ainsworth2023gitrebasinmergingmodels,Goodfellow-et-al-2016}), though to our knowledge this is the first time it has been explicitly accommodated in such a field-theoretic description.\footnote{For other physics-inspired approaches to symmetries in neural networks, see for example \cite{cohen2016groupequivariantconvolutionalnetworks,cheng2019covariancephysicsconvolutionalneural,iqbal2026spontaneoussymmetrybreakinggoldstone}.} As we will explain, each layer of the network can be thought of as a lattice site, at which sits an $N$-component field transforming in the fundamental representation. The role of the gauge fields living on the links is then played by the weight matrices, which transform in the adjoint. The result is a relatively intuitive description of the network as a 1-dimensional lattice of fields (layers), with interactions mediated by gauge fields (weight matrices), such that the action of the theory is invariant under local permutations. Let us here emphasize that in the present work, we confine ourselves to random neural networks at initialization, as in \cite{Grosvenor:2021eol} and similar theoretical studies, e.g., \cite{schoenholz2017deep}; we comment on the potential incorporation of training dynamics in context in sections \ref{sec:reloaded} and \ref{sec:discussion}.

The remainder of this paper is organized as follows: in sec. \ref{sec:reloaded}, we discuss the modifications to the NN/QFT duality \cite{Grosvenor:2021eol} to incorporate the layerwise permutation symmetry of the network. We construct the corresponding permutation-invariant action, and derive the bare neuron-neuron propagators and Feynman rules. In sec. \ref{sec:perturb}, we compute the leading perturbative corrections to the bare propagator in the $1/N$ expansion via a powerful recursive expression that incorporates infinitely-many Feynman diagrams. In sec. \ref{sec:scatter}, we explore the behaviour of higher-point amplitudes in the theory, and discuss their predicted scaling in $1/N$. We close in \ref{sec:discussion} with some discussion and directions for future work. In the appendices, we have included some details about the normalization of the partition function in appx. \ref{sec:norm}, and a reorganization of the infinite series for the exact two-point function, including a study of the pole structure arising from the zeta function expansion, in appx. \ref{sec:valiant}. Lastly, we focus in the present work on the theoretical development of the dual field theory; empirical explorations of the resulting model are deferred to future work \cite{TBA}.

\section{The DNN action revisited: permutation symmetry}\label{sec:reloaded}

In the course of obtaining the self-averaged partition function for DNNs in \cite{Grosvenor:2021eol}, the weights were assumed to be time- (i.e., layer-) independent. This is correct in the case of an RNN due to weight sharing, but is unrealistic for DNNs -- by which we mean multilayer perceptrons (MLPs), to which we restrict attention in this work -- whose weights are generally initialized independently at each layer. Furthermore, these networks enjoy layerwise permutation symmetry, whereby the activation functions at each layer can be acted upon by a permutation matrix, provided the weight matrix is acted upon by the inverse; see for example \cite{ainsworth2023gitrebasinmergingmodels,zhang2025permutationsymmetrytransformersrole}. Explicitly, let $h_\ell$ denote the vector of pre-activations at layer $\ell\in[0,L]$ (henceforth referred to as \emph{neurons}). An $L$-layer MLP is constructed as
\begin{equation}
	h_{\ell+1}=W_{\ell+1}\,\phi(h_{\ell})+b_{\ell+1}~,
	\qquad\quad
	h_0=x~,
	\label{eq:MLP}
\end{equation}
where $W_\ell$ is the weight matrix, $b_\ell$ is the vector of biases, and $x$ is the vector of initial data.\footnote{Note that some works instead express the network in terms of post-activations $y_\ell\coloneqq\phi(h_\ell)$, in which case the weight matrix appears inside the activation function, i.e., $y_{\ell+1}=\phi(W_{\ell+1}y_\ell+b_{\ell+1})$, but that convention is unwieldy for our purposes.} We assume for simplicity that all layers have width $N$, and are here suppressing the neuron indices for compactness.\footnote{As explained in the original work \cite{Grosvenor:2021eol}, the salient feature is that all layer widths go to infinity at the same rate, and thus we expect our results to hold even for networks with $N_\ell\neq N_\rho$ for $\ell\neq\rho$ provided that $N_\ell/N_\rho\to 1$ as the number of neurons per layer is taken to $\infty$.} We then observe that we can insert the identity in the form $\mathbb{1}=P^\mt P$, where $P\in S_N$ is a permutation matrix and $S_N$ is the finite symmetric group over $N$ elements,
\begin{equation}
	\begin{aligned}
		h_{\ell+1}&=W_{\ell+1}P_{\ell}^\mt P_{\ell}\,\phi(h_{\ell})+b_{\ell+1}=W_{\ell+1}P_{\ell}^\mt \phi\left(P_{\ell}h_{\ell}\right)+b_{\ell+1}\\
			  &=W_{\ell+1}P_{\ell}^\mt \phi\left(P_{\ell}W_{\ell}\,\phi(h_{\ell-1})+P_{\ell}\,b_{\ell}\right)+b_{\ell+1}~,
	\end{aligned}
\end{equation}
Observing that we can do the same at layer $\ell$, i.e., $W_\ell\,\phi\mapsto W_\ell P_{\ell-1}^\mt P_{\ell-1} \phi$ in the second line above, we see that the symmetry transformation for the network under $S_N$ is
\begin{equation}
	h_\ell\mapsto P_{\ell}\,h_\ell~,\qquad
	b_\ell\mapsto P_{\ell}\,b_\ell~,\qquad
	W_\ell\mapsto P_{\ell} W_\ell P_{\ell-1}^\mt~.
	\label{eq:sym}
\end{equation}
Note in particular that this is a local transformation: the permutation matrix may be chosen differently at each layer. Additionally, as we comment on more fully in sec. \ref{sec:discussion}, we note in passing that for linear networks, this can be elevated to a continuous rotation symmetry, but that the non-linearity spoils this since $R\phi(h)\neq \phi(Rh)$ for a continuous rotation $R$. In contrast $P\in S_N$ merely permutes the neuron indices, and hence commutes with the elementwise non-linearity $\phi$. 

Now, if $P$ were an element of a Lie group, we would proceed as in \cite{Grosvenor:2021eol}, taking the continuum limit (i.e., $L\to\infty$) in order to obtain a statistical field theory in $0\!+\!1$ dimensions, whereupon we could attempt to incorporate the symmetry in the usual language of gauge theory. However, since $S_N$ is discrete group, the standard notion of a connection on a principle bundle cannot be used to define the covariant derivative. This suggests that we refrain from taking the continuum limit, and instead treat the network as a $(0\!+\!1)$-dimensional lattice, with the discrete transformation $P$ living on the links. Indeed, the ability to treat discrete groups is one of the advantages of lattice gauge theory; see for example \cite{Esposito:2024uzw,PhysRevD.11.2098} and references therein, or \cite{Creutz_2023} for a pedagogical reference. Unlike higher-dimensional examples however, in the present case there is no plaquette term for the gauge field, since there is no (discrete) notion of parallel transport in $(0\!+\!1)$ dimensions. Physically, this corresponds to the fact that there are no equations of motion for the weight matrices\footnote{Incorporating training dynamics is expected to change this picture, since the update rule for stochastic gradient descent (SGD) acts like an external driving term. This adds a temporal dimension to the theory, which becomes $(1\!+\!1)$-dimensional, thus avoiding the geometrical constraint above. We hope to return to this potential extension in the future.} (which, in a standard gauge theory, would be induced by the field strength term corresponding to the 2-curvature in the bundle), which are instead drawn randomly at each lattice site (i.e., layer). Thus, the only point at which any gauge field enters the action is via the discrete covariant derivative in the matter kinetic term.

With the above in mind, let us revisit the construction of the field theory in \cite{Grosvenor:2021eol}, and see where we must deviate in order to incorporate the local symmetry \eqref{eq:sym}. Since we restrict our attention to MLPs -- the structure of which has neither stochasticity nor recurrence -- the stochastic differential equation that serves as the starting point for the description is simplified to the ordinary differential equation
\begin{equation}
	\mathrm{d} h=f(h)\mathrm{d}\ell~,
	\qquad\quad
	f(h)=W\phi(h)+b~,
	\label{eq:diffeq}
\end{equation}
cf. eq. (2.1) of \cite{Grosvenor:2021eol} (n.b., all equation references to that work shall refer to arXiv version 2.0). \emph{A priori}, ignoring any symmetries, we would interpret this in the It\^o discretization as
\begin{equation}
	h_{\ell}-h_{\ell-1}=f(h_{\ell-1})a~,
	\label{eq:noncovar}
\end{equation}
where $a\coloneqq \mathrm{d}\ell$ is the lattice spacing, representing the separation between layers. However, this must be modified in the presence of the permutation symmetry, since it does not result in a covariant derivative: the two pre-activations on the left-hand side of \eqref{eq:noncovar} transform under different permutation matrices. In standard lattice gauge theory, we would remedy this by inserting a link variable $U$ transforming in the adjoint representation to pushforward $h_{\ell-1}$ so that both terms are defined at the same lattice site $\ell$. Observe however that in the present system, this is formally the role played by the weight matrices $W$, cf. \eqref{eq:sym}. At a technical level, there is a minor difference in that $W$, being drawn from some Gaussian ensemble, is not an element of any representation of the permutation group $S_N$. Nonetheless, the action of $S_N$ on $W$ does not move one beyond the space of Gaussian matrices. Hence, we may interpret the weight matrices as the link variables connecting the lattice sites,\footnote{To be clear: each layer of $N$ neurons corresponds to a single lattice site, at which sits an $N$-component field $h_\ell$. The weight matrices then act like link variables between sites, which is physically natural insofar as this is precisely their role within the network, cf. \eqref{eq:MLP}. Incidentally, note that our covariant derivative requires the opposite of the usual convention in which the link variable $U_{\ell-1}$ acts to pullback site $\ell$ to site $\ell\!-\!1$; in that case $W_\ell$ would need to transform as $W_\ell\mapsto P_\ell W_\ell P_{\ell+1}^\mt$, in contrast to \eqref{eq:sym}.} and define the discrete covariant derivative via
\begin{equation}
	h_{\ell}-W_{\ell} h_{\ell-1}=f(h_{\ell-1}) a~,
	\label{eq:covito}
\end{equation}
so that all terms in this expression become right-multiplied by $P_\ell$ under the transformation \eqref{eq:sym}. We thus have the discrete covariant derivative
\begin{equation}
	D_\ell h\coloneqq\frac{ h_{\ell}- W_{\ell}h_{\ell-1}}{a}~,
	\qquad\textrm{with}\qquad
	D_\ell h\mapsto P_\ell D_\ell h
\end{equation}
under the transformation \eqref{eq:sym}.

We can now construct the path integral along similar lines as before \cite{Grosvenor:2021eol}. First, we express the probability of the state of the network at layer $L$ given some initial input vector $h_0=x$, where $x$ denotes the data, as a conditional probability marginalized over all internal (hidden) layers:
\begin{equation}
	p(h_L|x)=\int\prod_{\ell=1}^{L-1}\!\mathrm{d} h_\ell\,p(h_L|h_{\ell-1})p(h_{L-1}|h_{L-2})\ldots p(h_1|x)~.
	\label{eq:prob1}
\end{equation}
Since the network structure is deterministic, 
\begin{equation}
	p(h_\ell|h_{\ell-1})=\delta\left(h_\ell-y_\ell(h_{\ell-1})\right)~,
\end{equation}
where $h_\ell=y_\ell$ is the solution to \eqref{eq:covito},
\begin{equation}
	y_\ell(h_{\ell-1})=W_\ell h_{\ell-1}+f_{\ell-1}a~,\qquad \ell>0~,
\end{equation}
where we have adopted the shorthand $f_{\ell-1}\coloneqq f(h_{\ell-1})$. In other words, \eqref{eq:prob1} integrates over all possible internal network states that satisfy the covariantized network structure equation \eqref{eq:covito}. We then put this in a more useful form by expressing the delta functions in Fourier space,
\begin{equation}
	\delta(h)=\int_{-\infty}^\infty\!\frac{\mathrm{d}\omega}{2\pi}e^{i\omega^\mt h}
	=\int_{-i\infty}^{i\infty}\frac{\mathrm{d}z}{2\pi i}e^{z^\mt h}~,
	\label{eq:delfunc}
\end{equation}
where we have introduced the \emph{response field} $z=i\omega\in\mathbb{C}^N$ (see \cite{Grosvenor:2021eol,helias2019statistical} for an explanation of this terminology). Note that in order to preserve the permutation symmetry, $z$ must transform in the same manner as $h$, i.e., $z_\ell\mapsto P_\ell z_\ell$ under \eqref{eq:sym}. We thus obtain an integral over both neurons $h$ and response variables $z$,
\begin{equation}
	p(h_L|x)=\int\!\prod_{\ell=1}^{L-1}\!\mathrm{d}h_\ell\int_{-i\infty}^{i\infty}\!\prod_{\ell=1}^{L-1}\left(\frac{\mathrm{d}z_\ell}{2\pi i}\,e^{z_\ell^\mt(h_\ell-y_\ell)}\right)~.
	\label{eq:bound}
\end{equation}
Upon adding source terms $j_\ell$ and integrating over the final layer $L$,\footnote{Here $h_L$ is not the output layer -- whose width is typically fixed to $O(1)$ or $O(10)$ -- but the final hidden layer such that its width tends to $N\to\infty$. See appendix \ref{sec:norm} for comments on the normalization and interpretation of this conditional partition function $Z[j|x]$.} we obtain the path integral over all possible network states, conditioned on the data $x$:
\begin{equation}
	\begin{aligned}
		Z[j|x]
		&=\prod_{\ell=1}^{L}\int\!\mathrm{d}h_\ell\int_{-i\infty}^{i\infty}\!\frac{\mathrm{d}z_\ell}{2\pi i}\,e^{z_\ell^\mt(h_\ell-y_\ell)+j_\ell h_\ell a}\\
		&=\prod_{\ell=1}^{L}\int\!\mathrm{d}h_\ell\int_{-i\infty}^{i\infty}\!\frac{\mathrm{d}z_\ell}{2\pi i}\,\exp\left[z_\ell^\mt \left(h_\ell-W_\ell h_{\ell-1}-f_{\ell-1}a\right)+j_\ell h_\ell a\right]\\
		&=\prod_{\ell=1}^{L}\left\{\mathrm{d}h_\ell\int_{-i\infty}^{i\infty}\!\frac{\mathrm{d}z_\ell}{2\pi i}\right\}\exp\sum_{\ell=1}^L\left[z_\ell^\mt\left(h_\ell-W_\ell h_{\ell-1}-f_{\ell-1}a\right)+j_\ell h_\ell a\right]~.
	\end{aligned}
\end{equation}
Now, rather than take the continuum $a\to0$ limit for the reasons mentioned above, we proceed directly to the self-averaging discussed in \cite{Grosvenor:2021eol}: that is, we are interested in random neural networks at initialization, in which the trainable parameters $W$, $b$ are independent Gaussian variables,
\begin{equation}
	W^{ij}\sim\mathcal{N}(0,\sigma_w^2/N)~,
	\qquad\qquad
	b^i\sim\mathcal{N}(0,\sigma_b^2)~,
	\label{eq:normal}
\end{equation}
where the normalization by $1/N$ for the weights is needed to insure that the activations remain $O(1)$. After integrating over these parameters, the partition function will describe an ensemble of such networks, where the observables of interest are computed in the ensemble average $\left<Z[j|x]\right>_{W,b}\eqqcolon \overline{Z}[j|x]$ at large $N$; see \cite{Grosvenor:2021eol,helias2019statistical} for further explanation on this point. The ensemble average partition function is then
\begin{equation}
	\overline{Z}[j|x]=\prod_{\ell=1}^{L}\left\{\int\!\mathrm{d}h_\ell\int_{-i\infty}^{i\infty}\!\frac{\mathrm{d}z_\ell}{2\pi i}\int\!\mathrm{d}W_\ell\,\rho(W_\ell)\!\int\!\mathrm{d}b_\ell\,\rho(b_\ell)\!\right\}\exp\sum_{\ell=1}^L\left[z_\ell^\mt\left(h_\ell-W_\ell h_{\ell-1}-f_{\ell-1}a\right)+j_\ell h_\ell a\right]~,
\end{equation}
where $\rho(W)$, $\rho(b)$ are Gaussian probability density functions with the moments specified in \eqref{eq:normal}. In a normal (lattice) gauge theory, we would proceed to add a field strength term for the link variables -- the role of which is here being played by $W$ -- in order to self-consistently generate their equations of motion. As mentioned above however, there is no field strength in $0\!+\!1$ dimensions, since all plaquette terms vanish (in the continuum case for example, parallel transport around any closed path in the bundle would be identically zero, since one can only move forwards and backwards along a line). Thus, despite randomly fluctuating at each link, the weight fields for networks at initialization are non-dynamical in the sense that there are no equations of motion governing their evolution. Hence, thanks to the simple Gaussian measures, we can directly perform the integrals over the weights and biases, in effect integrating-out the link variables. For the weights, we have
\begin{equation}
	\begin{aligned}
		{}&\int\!\prod_{\ell=1}^{L}\sqrt{\tfrac{N}{2\pi\sigma_w^2}}\,\mathrm{d}W_\ell\exp\sum_{\ell=1}^L\left\{-\frac{N}{2\sigma_w^2}W_\ell^2-z_\ell^\mt\left(\phi(h_{\ell-1})a+h_{\ell-1}\right)W_\ell\right\}\\
		{}&=\exp\left\{\frac{\sigma_w^2}{2N}\sum_{\ell=1}^Lz_\ell^2\left(\phi(h_{\ell-1})a+h_{\ell-1}\right)^2\right\}~,
	\end{aligned}
\end{equation}
cf. (2.22) of \cite{Grosvenor:2021eol}, which holds for a general RNN. 
In contrast, here -- in addition to the discreteness -- the expression is local rather than bi-local in the depth, as a consequence of treating the weight matrices (i.e., link variables) $W_\ell$ as layer-dependent. Similarly, for the biases,
\begin{equation}
	\begin{aligned}
		\int\!\prod_{\ell=1}^{L}\frac{\mathrm{d}b_\ell}{\sqrt{2\pi\sigma_b^2}}\exp\sum_{\ell=1}^L\left[-\frac{b_\ell^2}{2\sigma_b^2}-a z_\ell^\mt b_\ell \right]
		=\exp\left\{\frac{a^2\sigma_b^2}{2}\sum_{\ell=1}^Lz_\ell^2\right\}~.
	\end{aligned}
\end{equation}
Ignoring the sources, our partition function then reads
\begin{equation}
	\overline{Z}=\prod_{\ell=1}^{L}\left\{\int\!\mathrm{d}h_\ell\int_{-i\infty}^{i\infty}\!\frac{\mathrm{d}z_\ell}{2\pi i}\right\}\exp\sum_{\ell=1}^L\sum_{i=1}^N\left\{z_\ell^i h_\ell^i+\frac{1}{2}\left(z_\ell^i\right)^2\left[\frac{\sigma_w^2}{N}\sum_{j=1}^N\left(\phi(h_{\ell-1}^j)a+h_{\ell-1}^j\right)^2+a^2\sigma_b^2\right]\right\}~,
	\label{eq:Z2}
\end{equation}
where we have written the dot products of vectors explicitly in terms of sums over neuron indices $i,j\in[1,N]$ to make clear the following observation: the system has decoupled into $N$ independent response fields $z_\ell^i$ with identical couplings given by $(a\sigma_b)^2$ plus the sum over $j$. This motivates the introduction of the local, permutation-invariant object
\begin{equation}
	\frak{A}_\ell\coloneqq\sqrt{\frac{\sigma_w^2}{N}}\sum_{j=1}^N\left(\phi(h_{\ell-1}^j)a+h_{\ell-1}^j\right)^2~,
	\label{eq:aux}
\end{equation}
where again \cite{Grosvenor:2021eol} $\sigma_w^2$ plays the role of the 't Hooft coupling. The benefit of this is that it facilitates the large-$N$ analysis by allowing us to represent the collective influence of (some function of) neurons $h$ in terms of an integral over the new fluctuating field $\frak{A}$, which we constrain to \eqref{eq:aux} by inserting a delta functional exactly as we did in \eqref{eq:delfunc}; i.e., we write \eqref{eq:Z2} in the form
\begin{equation}
	\begin{aligned}
		\overline{Z}=\prod_{\ell=1}^{L}\left\{\int\!\mathrm{d}h_\ell\int_{-i\infty}^{i\infty}\!\frac{\mathrm{d}z_\ell}{2\pi i}\int\!\mathrm{d}\frak{A}_\ell\right\}\exp\sum_{\ell=1}^L&\sum_{i=1}^N\left\{z_\ell^i h_\ell^i+\frac{1}{2}\left(z_\ell^i\right)^2\left(\sqrt{\frac{\sigma_w^2}{N}}\frak{A}_\ell+a^2\sigma_b^2\right)\right\}\\
		\times&\,\delta\Bigg(\frak{A}_\ell-\sqrt{\frac{\sigma_w^2}{N}}\sum_{j=1}^N\left(\phi(h_{\ell-1}^j)a+h_{\ell-1}^j\right)^2\Bigg)\\
			  =\prod_{\ell=1}^{L}\left\{\int\!\mathrm{d}h_\ell\mathrm{d}\frak{A}_\ell\int_{-i\infty}^{i\infty}\!\frac{\mathrm{d}z_\ell \mathrm{d}A_\ell}{2\pi i}\right\}\exp\sum_{\ell=1}^L
						 &\left\{\sum_{i=1}^N\left[z_\ell^i h_\ell^i+\frac{1}{2}\left(z_\ell^i\right)^2\left(\sqrt{\frac{\sigma_w^2}{N}}\frak{A}_\ell+a^2\sigma_b^2\right)\right]\right.\\
					  +A_\ell\!&\left.\left[\frak{A}_\ell-\sqrt{\frac{\sigma_w^2}{N}}\sum_{j=1}^N\left(\phi(h_{\ell-1}^j)a+h_{\ell-1}^j\right)^2\right]\right\}~,
	\end{aligned}
	\label{eq:Spre}
\end{equation}
where $A_\ell$ is another complex auxiliary field, analogous to $z_\ell$, with the important difference that $A_\ell$ is a scalar while $z_\ell$ is a vector. To help organize this, we group all quadratic terms into the free part of the action,
\begin{equation}
	S_0\coloneqq\sum_{\ell=1}^L\left[\sum_{i=1}^Nz_\ell^i\left(h_\ell^i+\frac{1}{2}a^2\sigma_b^2\,z_\ell^i\right)+A_\ell\frak{A}_\ell\right]~,
\end{equation}
and collect the remaining terms in the interacting part,
\begin{equation}
	S_\mathrm{int}\coloneqq\sqrt{\frac{\sigma_w^2}{N}}\sum_{\ell=1}^L\sum_{i=1}^N\left[\frac{1}{2}\frak{A}_\ell\left(z_\ell^i\right)^2-A_\ell\left(a\,\phi(h_{\ell-1}^i)+h_{\ell-1}^i\right)^2\right]~.
\end{equation}
Incidentally, observe that in the large-$N$ limit, the interactions vanish, and we recover a free theory as expected.

Before proceeding to solve for the propagators, we first perform any necessary field redefinitions in order to ensure that we are expanding around the true vacuum state of the theory. To that end, we obtain the vacuum expectation value (vev) for each of the fields appearing in the action by solving the corresponding equations of motion:
\begin{equation}
	\begin{aligned}
		\frac{\delta\ln \overline{Z}}{\delta h_\rho^j}=0&\implies
		\left<z_\rho^j\right>=2\sqrt{\frac{\sigma_w^2}{N}}\left<A_{\rho+1}\left(a\phi(h_{\rho+1}^j)+h_{\rho+1}^j\right)\left(a\phi'(h_{\rho+1}^j)+1\right)\right>~,\\
		\frac{\delta\ln \overline{Z}}{\delta z_\rho^j}=0&\implies
		\left<h_\rho^j\right>=-\left<\left(a^2\sigma_b^2+\sqrt\frac{\sigma_w^2}{N}\frak{A}_\rho\right)z_\rho^j\right>~,\\
		\frac{\delta\ln \overline{Z}}{\delta A_\rho^j}=0&\implies
	\left<\frak{A}_\rho\right>=\sqrt{\frac{\sigma_w^2}{N}}\left<\sum_{i=1}^N\left(a\phi(h_{\rho-1}^i)+h_{\rho-1}^i\right)^2\right>~,\\
		\frac{\delta\ln \overline{Z}}{\delta \frak{A}_\rho^j}=0&\implies
		\left<A_\rho\right>=-\frac{1}{2}\sqrt{\frac{\sigma_w^2}{N}}\left<(z_\rho^i)^2\right>~.
	\end{aligned}
\end{equation}
However, all $n$-point functions of the response field $z$ vanish \cite{Grosvenor:2021eol,helias2019statistical}. Additionally, the lack of any mixed quadratic terms in the action -- that is, those involving either $A_\ell$ or $\frak{A}_\ell$ with any other field -- implies that the remaining correlators factorize, thus reducing the above to\footnote{Note that $\phi$ is not a fundamental field, so we do not specify a vev for either this or $\phi'$. We will address this issue momentarily.}
\begin{equation}
	\begin{aligned}
		\left<z_\rho^j\right>&=0\\
		\left<h_\rho^j\right>&=-\sqrt\frac{\sigma_w^2}{N}\left<\frak{A}_\rho \right>\left<z_\rho^j\right>=0\\
		\left<\frak{A}_\rho\right>&=\sqrt{\frac{\sigma_w^2}{N}}\left<\sum_{i=1}^N\left(a\phi(h_{\rho-1}^i)+h_{\rho-1}^i\right)^2\right>\eqqcolon c_{\rho-1}\\
		\left<A_\rho\right>&=0~,
	\end{aligned}
	\label{eq:vevs}
\end{equation}
and thus we see that the only non-zero vev is that associated to the field $\frak{A}_\rho$. Hence, we perform the field redefinition
\begin{equation}
	\frak{A}_\rho\mapsto \frak{A}_\rho+c_{\rho-1}
\end{equation}
in the action $S_0+S_\mathrm{int}$.\footnote{That is, we define a new field $\frak{A}_\rho'\coloneqq \frak{A}_\rho-c_{\rho-1}$ so that $\left<\frak{A}_\rho'\right>=0$, rewrite the action in terms of $\frak{A}_\rho'$, and then drop the primes for compactness.} Expressed around the true vacuum state, our theory is then
\begin{equation}
	\overline{Z}=\prod_{\ell=1}^{L}\left\{\int\!\mathrm{d}h_\ell\mathrm{d}\frak{A}_\ell\int_{-i\infty}^{i\infty}\!\frac{\mathrm{d}z_\ell \mathrm{d}A_\ell}{2\pi i}\right\}\,e^{S_0+S_\mathrm{int}}~,
\end{equation}
where the coefficient of the $z^2$ term in the free part of the action picks up a contribution from the non-zero vev above,
\begin{equation}
	S_0\coloneqq\sum_{\ell=1}^L\left\{\sum_{i=1}^N\left[z_\ell^ih_\ell^i+\frac{1}{2}\left(a^2\sigma_b^2+\sqrt{\frac{\sigma_w^2}{N}}c_{\ell-1}\right)\left(z_\ell^i\right)^2\right]+A_\ell\frak{A}_\ell\right\}~,
	\label{eq:S0}
\end{equation}
while the interacting part remains unchanged (but which we repeat here for convenience),
\begin{equation}
	S_\mathrm{int}=\sqrt{\frac{\sigma_w^2}{N}}\sum_{\ell=1}^L\sum_{i=1}^N\left[\frac{1}{2}\frak{A}_\ell\left(z_\ell^i\right)^2-A_\ell\left(a\,\phi(h_{\ell-1}^i)+h_{\ell-1}^i\right)^2\right]~.
	\label{eq:SintnoTaylor}
\end{equation}
There is one final step we must perform before the theory above becomes tractable: to apply standard field-theoretic methods, the non-linearity $\phi(h)$ must be represented as a polynomial of the fundamental field $h$. For technical reasons \cite{Roberts:2021fes}, we will assume that $\phi$ admits an expansion near $h=0$ of the form 
\begin{equation}
	\phi(h)=\sum_{n=1}^\infty\alpha_nh^n~.
	\label{eq:Taylor}
\end{equation}
Importantly, note that we are assuming that the activation function is analytic at the origin,\footnote{While the requirement of differentiability excludes rectified linear units (ReLU), the series \eqref{eq:Taylor} can accommodate a range of other popular activations functions, including tanh and swish (also known as sigmoid linear units (SiLU)). That the function be differentiable at the origin is desired on theoretical grounds \cite{Roberts:2021fes}, but since the set of neurons which lie exactly at this point is measure zero, it is unclear how stringent this condition is in practice, which may partially explain why popular functions such as ReLU perform better than such an analysis would expect. We thank David Berman for discussion on this point.} but that we do \emph{not} assume a finite radius of convergence for this expansion, since this requires imposing extra conditions on the preactivations that are not generically met in practice (at least not without specific interventions such as batch norm \cite{ioffe2015batchnormalizationacceleratingdeep}). We note however that truncating the Taylor series regardless may not be a bad approximation in practice. Taking $\phi\coloneqq\tanh$ as a prototypical example, analytical convergence of the expansion \eqref{eq:Taylor} requires $|h|\!<\!\pi/2$, whereas the average standard deviation for the preactivations in a network with $L\!=\!100$ initialized near the tree-level critical point ($\sigma_w^2=1.76$, $\sigma_b^2=0.05$) is approximately $\pi/4$, meaning that only about $5\%$ of preactivations fall outside this range. 
Below, we will see that the infinite series of Feynman diagrams, at least at leading order in this expansion, can nonetheless be managed by treating \eqref{eq:Taylor} as a formal power series.
Generically, this leads to interaction terms of the form
\begin{equation}
	A_\ell\left(a\,\phi(h_{\ell-1}^i)+h_{\ell-1}^i\right)^2
	=A_\ell\left[a^2\sum_{n,m=1}^\infty\alpha_n\alpha_m(h_{\ell-1}^i)^n(h_{\ell-1}^i)^m+2ah_{\ell-1}^i\sum_{n=1}^\infty\alpha_n(h_{\ell-1}^i)^n+(h_{\ell-1}^i)^2
	\right]~.
	\label{eq:formalinteract}
\end{equation}
We note that each term is still permutation invariant, since in index notation, this amounts to the statement that it does not matter in which order we perform the sum over neural indices $i\in[1,N]$.

As a concrete application of this framework, let us specify to the case $\phi=\tanh$, since as alluded above this is the prototypical activation function for such theoretical studies, e.g., \cite{schoenholz2017deep,poole2016exponential,Grosvenor:2021eol}. We remind the reader that the corresponding Taylor expansion \eqref{eq:Taylor} is then
\begin{equation}
	\tanh(h)=\sum_{n=1}^\infty f_nh^n~,
	\label{eq:tanhexp}
\end{equation}
where for future convenience we have denoted the coefficients
\begin{equation}
	f_n=\frac{2^{2n}\left(2^{2n}-1\right)B_{2n}}{(2n)!}~,
	\qquad \forall n\in\mathbb{Z}^+~,
	\label{eq:fcoef}
\end{equation}
where $B_{k}$ is the $k^\mathrm{th}$ Bernoulli number. In this case the interaction part of the action \eqref{eq:SintnoTaylor} may be written
\begin{equation}
	S_\mathrm{int}=\sqrt{\frac{\sigma_w^2}{N}}\sum_{\ell=1}^L\sum_{i=1}^N\left[\frac{1}{2}\frak{A}_\ell\left(z_\ell^i\right)^2-A_\ell\left((h_{\ell-1}^i)^2+\sum_{k\geq2}^{\mathrm{even}}g_k(h_{\ell-1}^i)^k\right)\right]~,
	\label{eq:Sintformal}
\end{equation}
where the sum over $k\geq2$ runs over all even integers, and the coefficients are given by
\begin{equation}
	g_k\coloneqq a^2c_k+2af_{k/2}~,
	\qquad \forall k\in 2\mathbb{Z}^+~,
	\label{eq:gcoef}
\end{equation}
with $f_n$ defined above, and 
\begin{equation}
	c_k=2^{k+2}\sum_{r\textrm{ odd}}^k\frac{\left(2^{r+1}-1\right)\left(2^{k-r+1}-1\right)B_{r+1}B_{k-r+1}}{(r+1)!(k-r+1)!}~,
	\qquad \forall k\in 2\mathbb{Z}^+~,
	\label{eq:ccoef}
\end{equation}
arising from the Cauchy product (from the $\phi^2$ quadratic term). Note that in these expressions, $k$ is even, $r\geq 1$ is odd, and $n$ runs over all positive integers. We will have more to say about these formal power series when computing corrections to the correlation function between neurons below, cf. subsec. \ref{sec:perturb} and appendix \ref{sec:valiant}. First however, we obtain the tree-level result by solving for the neuron-neuron propagator for the model. 

\subsection{Bare propagators}

Observing the form of $S_0$ \eqref{eq:S0}, we see that $h$ and $z$ cannot really be thought of as independent fields, since the first term allows them to freely propagate into each other; similarly for $\frak{A}$ and $A$. Hence, to solve for the propagators, we follow \cite{Grosvenor:2021eol} in introducing the two-component fields
\begin{equation}
	y_\ell\coloneqq
	\begin{pmatrix}
		h_\ell\\z_\ell
	\end{pmatrix}~,
	\qquad\qquad
	B_\ell\coloneqq
	\begin{pmatrix}
		\frak{A}_\ell\\A_\ell
	\end{pmatrix}~,
	\label{eq:twocomp}
\end{equation}
and rewrite the quadratic part of the action in the form\footnote{For clarity, note that the transpose operation on nested vectors is here defined to act on all levels, not just the outermost. That is, $y_\ell$ is understood as a column of column vectors, and $y_\ell^\mt\coloneqq(h_\ell^\mt,\,z_\ell^\mt)$ a row of row vectors, so that the $z_\ell$ and $h_\ell$ vectors combine to form scalars as above. If one is bothered by this shorthand, one can simply write the sums over contracted neuron indices explicitly.}
\begin{equation}
	S_0=-\frac{1}{2}\sum_{\ell,\rho=1}^L\left(y_\ell^\mt\,\Xi_{\ell\rho}\,y_\rho+B_\ell\Upsilon_{\ell\rho} B_\rho\right)~,
\end{equation}
where
\begin{equation}
	\Xi_{\ell\rho}\coloneqq
	\delta_{\ell\rho}
	\begin{pmatrix}
		0 & \; & -1 \\
		-1 & \; & -a^2\sigma_b^2-\sqrt{\frac{\sigma_w^2}{N}}c_{\ell-1}
	\end{pmatrix}~,
	\qquad\qquad
	\Upsilon_{\ell\rho}\coloneqq\delta_{\ell\rho}
	\begin{pmatrix}
		0 & \, & -1 \\ -1 & \, &  0
	\end{pmatrix}~.
\end{equation}
In the continuum, the propagators would be obtained as the Green functions for these operators. Here, we may obtain them analogously via the requirement that they act as matrix inverses, i.e., for $\Xi$ we have
\begin{equation}
	\sum_{\tau=1}^L\Xi_{\ell\tau}G_{\tau\rho}=\delta_{\ell\rho}\mathbb{1}~,
\end{equation}
which implies
\begin{equation}
	G_{\ell\rho}=
	\delta_{\ell\rho}
	\begin{pmatrix}
		a^2\sigma_b^2+\sqrt{\frac{\sigma_w^2}{N}}c_{\ell-1} & \, & -1 \\ 
		-1 & \, & 0
	\end{pmatrix}~.
	\label{eq:Green}
\end{equation}
Similarly, for $\Upsilon$ we have simply
\begin{equation}
	D_{\ell\rho}=\delta_{\ell\rho}
	\begin{pmatrix}
		0 & \, & -1 \\ -1 & \, &  0
	\end{pmatrix}~.
\end{equation}
Note that these are much simpler than the corresponding expressions for RNNs \cite{Grosvenor:2021eol} thanks to the discretized nature of the theory. In particular, the mixed propagators, given by the off-diagonal elements, are trivial, and merely impose the delta function constraint on the layers; hence: 
\begin{equation}
	\left<h_\ell^i z_\rho^j\right>=\left<z_\ell^i h_\rho^j\right>=-\delta_{\ell\rho}\delta^{ij}~,
\end{equation}
and similarly for $D_{\ell\rho}$,
\begin{equation}
	\left<A_\ell\frak{A}_\rho\right>=-{\delta_\ell\rho}~.
\end{equation}
However, the upper-left element of \eqref{eq:Green} yields a recursive expression for the neuron-neuron propagator, which is our primary interest:
\begin{equation}
	\left<h_\ell^i h_\rho^j\right>=\,\delta_{\ell\rho}\delta^{ij}\left(a^2\sigma_b^2+\frac{\sigma_w^2}{N}\left<\sum_{k=1}^N\left(a\phi(h_{\ell-1}^k)+h_{\ell-1}^k\right)^2\right>\right)\eqqcolon \delta_{\ell\rho}\delta^{ij}\Delta_\ell,
	\label{eq:recursivehh}
\end{equation}
where we have stripped-off the delta functions to denote the $\ell^\mathrm{th}$ layer auto-correlation ${\Delta_\ell\coloneqq\left<(h_\ell^i)^2\right>}$ on the far right-hand side, suppressing the neuron index for reasons which will be explained momentarily. Notably, these expressions predict that the interlayer correlation vanishes, which can be understood a consequence of gauge invariance: since the permutation symmetry \eqref{eq:sym} is local, only locally-invariant quantities -- such as $h_\ell^2$ or $h_\ell^2h_\rho^2$ -- are observables. Additionally, the lack of intralayer connections implies that the only coupling to other neurons is via previous layers, which is seen in the recursion relation here.

To unpack this recursive expression, we note that since the tree-level action is free (i.e., Gaussian), Wick's theorem implies that all higher-point correlators on the right-hand side -- cf. the general form of the series expansion in \eqref{eq:formalinteract} -- factorize into sums of products of two-point correlators. Again specifying to the case $\phi=\tanh$ for concreteness, the formal power series expansion of the squared quantity was given in \eqref{eq:Sintformal}; hence:
\begin{equation}
	\begin{aligned}
		\Delta_\ell&=\,a^2\sigma_b^2+\frac{\sigma_w^2}{N}\sum_{k=1}^N\left<(h_{\ell-1}^k)^2+\sum_{m\geq2}^{\mathrm{even}}g_m(h_{\ell-1}^i)^m\right>\\
			   &=\,a^2\sigma_b^2+\frac{\sigma_w^2}{N}\sum_{k=1}^N\sum_{m\geq2}^{\mathrm{even}}\gamma_m\,(m-1)!!\left<(h_{\ell-1}^k)^2\right>^{m/2}\\
	\end{aligned}
	\label{eq:hhrecpre}
\end{equation}
where for compactness we have defined
\begin{equation}
	\gamma_m\coloneqq
	\begin{cases}
		g_2+1~,\quad &m=2~,~\\
		g_m~,\quad &m>2~,
	\end{cases}
	\label{eq:gprime}
\end{equation}
with the coefficients $g_m$ given in \eqref{eq:gcoef}. We then recognize $\Delta_{\ell-1}$ on the right-hand side, but let us here pause to clarify a potential confusion. In the expression
\begin{equation}
	\frac{1}{N}\sum_{k=1}^N\left<(h_{\ell-1}^k)^2\right>^{m/2}
	=\frac{1}{N}\sum_{k=1}^N\left(\Delta_{\ell-1}\right)^{m/2}
	=\left(\Delta_{\ell-1}\right)^{m/2}~,
	\label{eq:ensens}
\end{equation}
the expectation value $\left<\ldots\right>$ denotes the ensemble average with respect to our partition function, i.e., the average over different network initializations (concretely, different seed values for the randomized selection of weights and biases). Meanwhile the $\tfrac{1}{N}\sum\nolimits_{i=1}^N$ computes the average over neurons in the given layer. However, since the neurons are all identical, the ensemble average $\left<(h_\ell^i)^2\right>\coloneqq\Delta_\ell$ given in \eqref{eq:hhrecpre} does not actually depend on the neuron index $i$, and hence any function of $\Delta_{\ell-1}$ may be moved outside the sum over $i$, as in the last equality in \eqref{eq:ensens}. The recursive expression \eqref{eq:hhrecpre} therefore becomes
\begin{equation}
	\Delta_\ell=\,a^2\sigma_b^2+\sigma_w^2\sum_{m\geq2}^{\mathrm{even}}\gamma_m\,(m-1)!!\,\Delta_{\ell-1}^{m/2}~.
	\label{eq:hhrec}
\end{equation}
The exception to the above -- for which the last equality in \eqref{eq:ensens} does not hold -- is the boundary condition $\Delta_0^i=\left<(h_0^i)^2\right>$ where $h_0^i=x^i$ is the data input to neuron $i$. Since there is no further recursion, the ensemble average must be computed for each fixed neuron -- i.e., each element of the data vector $x$ -- and then $(\Delta_0^i)^{m/2}$ is averaged over neurons (e.g., over the pixels in the specified input image if $x$ represents an element of the MNIST dataset). The expression \eqref{eq:hhrec} then allows us to compute the expected trajectory of (the auto-correlation of) a given data point through the ensemble of networks. From the field theory perspective, this expression says that the contribution from all lower layers is $O(1)$ in the perturbative expansion, which is consistent with the fact that this is a tree-level expression. We will consider perturbative corrections to this trajectory in sec. \ref{sec:perturb}, but first we must obtain the Feynman rules for our theory.

\subsection{Feynman rules}\label{sec:rules}

From the propagators above and the form of the interactions \eqref{eq:Sintformal}, we obtain the following Feynman rules for the theory. Recall that Feynman diagrams are read from left to right, but that within correlation functions, operators are ordered from right to left. Thus for example, a $zh$ propagator corresponds to $\left<h^\mt z\right>$. For clarity, we have referred to propagators via the corresponding correlation functions below. Following \cite{Grosvenor:2021eol}, we have kept the convention that all auxiliary fields ($z$ and $A$) are shaded grey, while all real neuron operators ($h$ and $\frak{A}$) are solid black.\\

\hspace{-1.1cm}\noindent\begin{minipage}{0.5\textwidth}
	\centering
	\textbf{Propagators:}
	\vspace{1em}
	\begin{description}
		\item[$\left<h_\ell^i h_\rho^j\right>$:] $\raisebox{-0.4\height}{\includegraphics[width=0.4\textwidth]{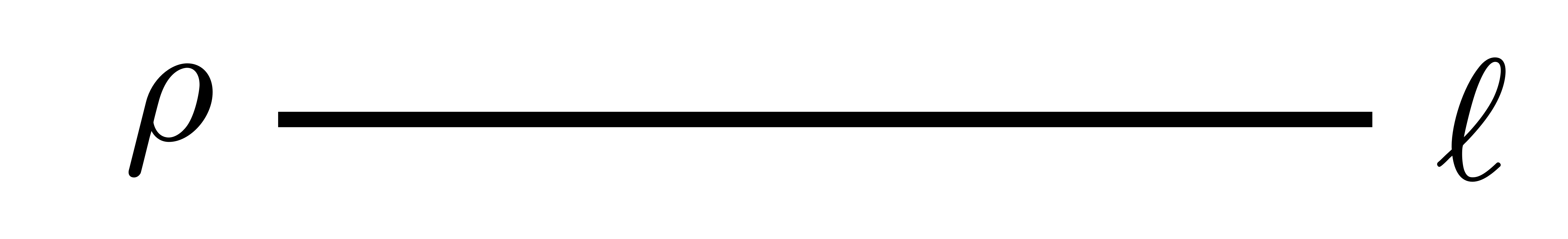}}=\delta_{\ell\rho}\delta^{ij}\Delta_\ell$\\
			\vspace{1em}
		\item[$\left<h_\ell^i z_\rho^j\right>$:] $\raisebox{-0.4\height}{\includegraphics[width=0.4\textwidth]{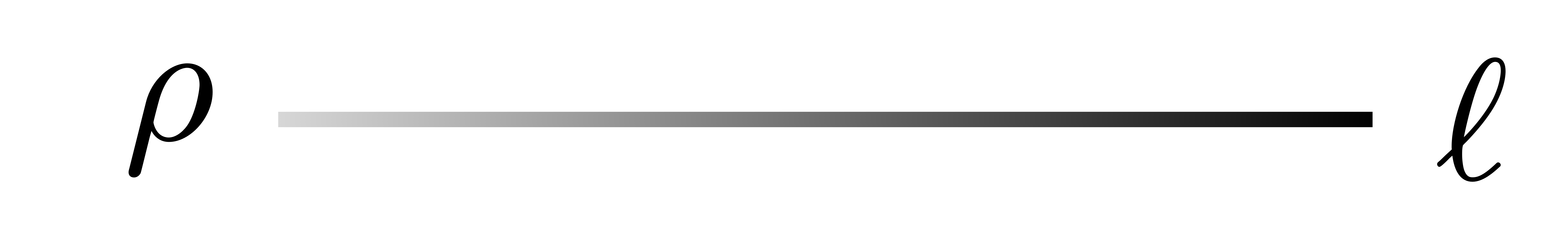}}=-\delta_{\ell\rho}\delta^{ij}$\\
			\vspace{1em}
		\item[$\left<z_\ell^i h_\rho^j\right>$:] $\raisebox{-0.4\height}{\includegraphics[width=0.4\textwidth]{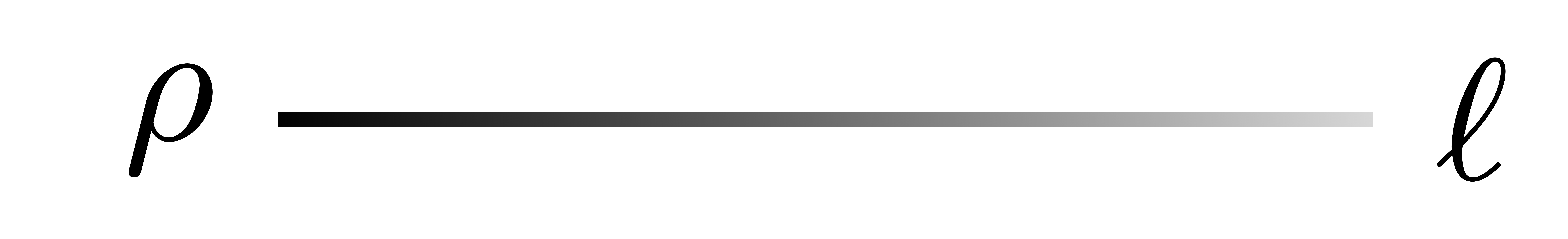}}=-\delta_{\ell\rho}\delta^{ij}$\\
			\vspace{1em}
		\item[$\left<\frak{A}_\ell A_\rho\right>$:] $\raisebox{-0.4\height}{\includegraphics[width=0.4\textwidth]{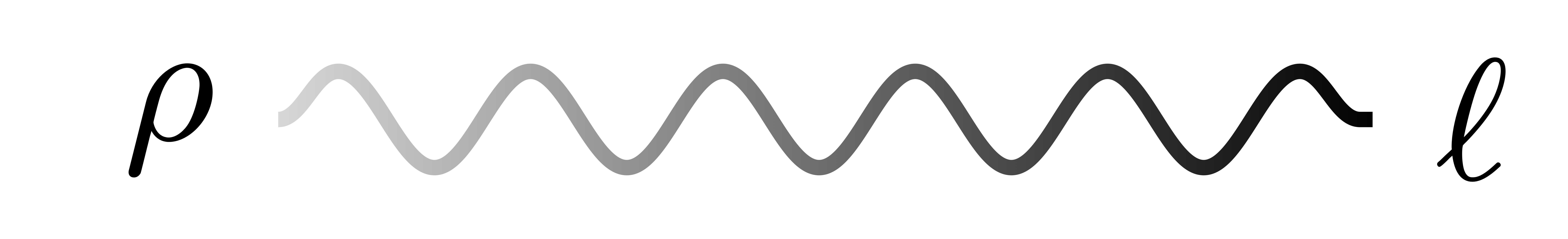}}=-\delta_{\ell\rho}$
	\end{description}
\end{minipage}
\hspace{0.05\textwidth}
\begin{minipage}{0.5\textwidth}
	\centering
	\textbf{Vertices:}
	\begin{description}
		\item[$\frak{A}_\ell(z_\ell^i)^2$:] $\quad\raisebox{-0.45\height}{\includegraphics[width=0.4\textwidth]{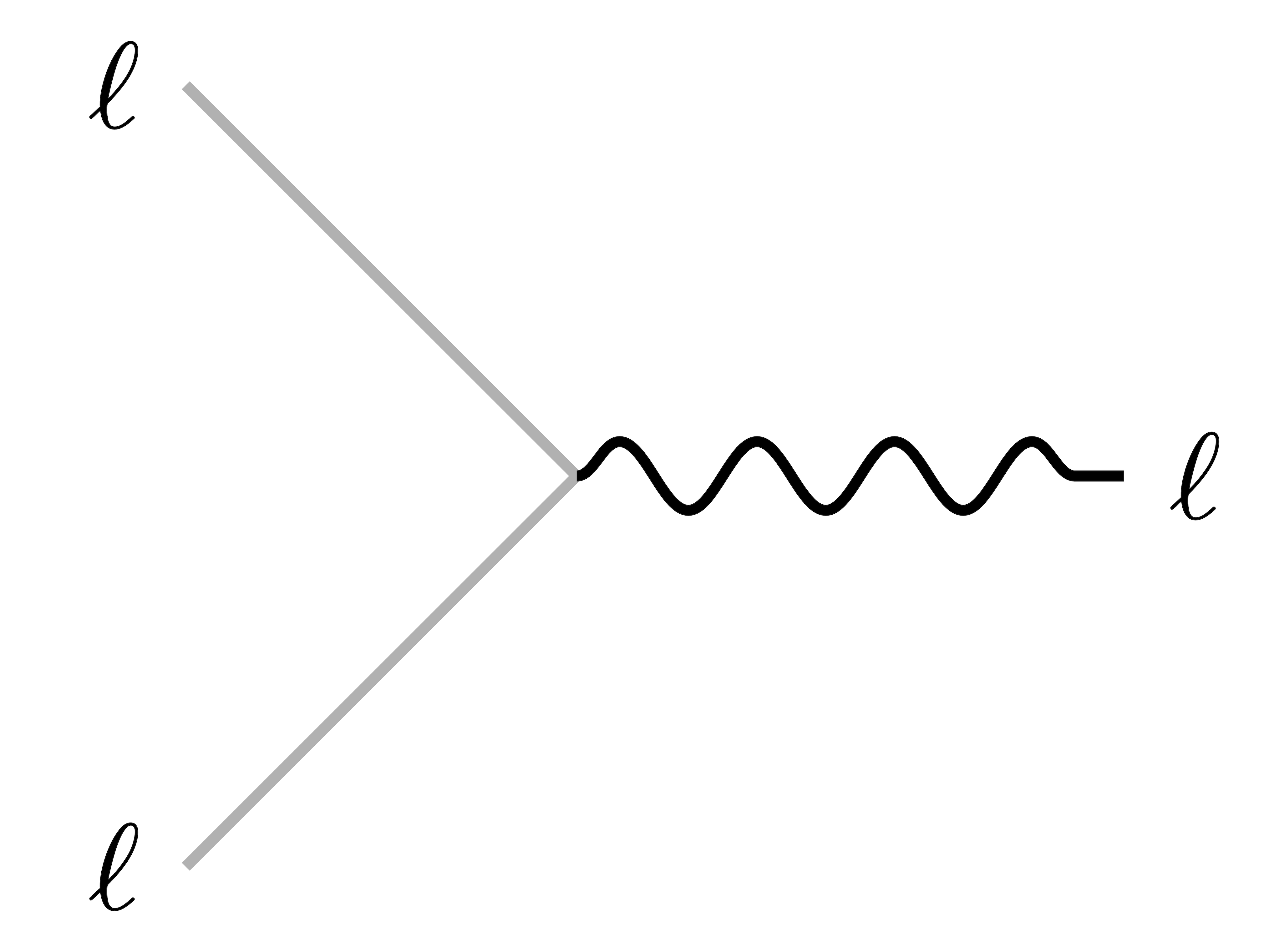}}=\frac{1}{2}\sqrt{\frac{\sigma_w^2}{N}}$
		\item[$A_\ell(h_{\ell-1}^i)^n$:] $\raisebox{-0.45\height}{\includegraphics[width=0.41\textwidth]{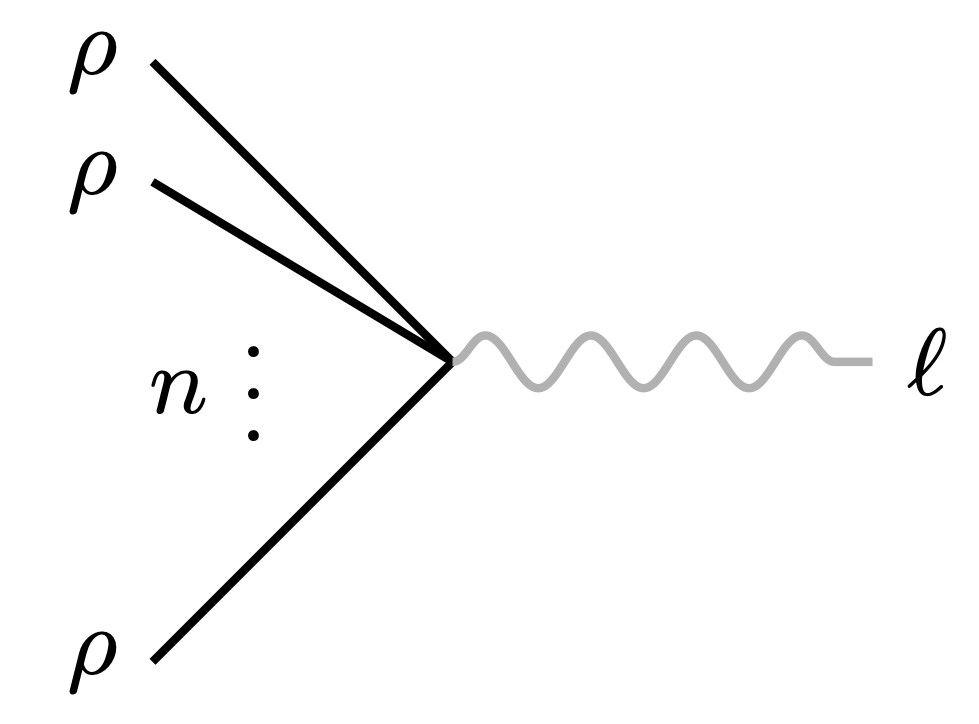}}{=-\gamma_n\sqrt{\frac{\sigma_w^2}{N}}\delta_{\rho,\ell-1}}$
	\end{description}
\end{minipage}\\

Note that here we have written the $Ah^n$ vertex using the coefficients for $\phi=\tanh$, with $\gamma_n$ given in \eqref{eq:gprime}, though abstractly the rules are the same for any polynomial function, cf. \eqref{eq:formalinteract}. Importantly, observe that this vertex couples two adjacent layers. 
Furthermore, the rules make it very easy to track the $N$-scaling of diagrams: \emph{de facto}, each $A\frak{A}$ propagator will carry a factor of $1/N$ (from the vertices at either end), while each loop carries a factor of $N$ (from the sum over contracted neuron indices). Additionally, the trivial nature of all non-$hh$ propagators makes the resulting Feynman diagrams quite simple, since they collapse to some number of vertex powers and factors of $N$ running in loops. Nonetheless, the fact that we are formally retaining all terms in the Taylor expansion of the non-linearity results in substantial richness in the diagrammatic expansion. In the next section, we will explore this by computing the quantum (i.e., statistical) corrections to the bare two-point function $\Delta_\ell$ from \underline{all} loop diagrams at leading order in $1/N$. Note that unlike most works in the machine learning literature, we find that this correction appears at $O(1)$, rather than $O(1/N)$. As explained in \cite{Grosvenor:2021eol}, these correspond to statistical fluctuations in the ensemble of networks, and persist even as one backs away from the infinite-width limit.

\section{Perturbative corrections to the neuron propagator}\label{sec:perturb}

As in \cite{Grosvenor:2021eol}, the leading corrections to the bare propagator $\Delta_\ell$ are $O(1)$, and are structurally reminiscent of the \emph{cactus diagrams} familiar from vector models, with $\sigma_w^2$ playing the role of the 't Hooft coupling. Unlike that work however, the simplicity of the lattice Feynman rules allows us to retain arbitrarily-many loops even at strong coupling, though as alluded above the empirical validity of the expansion may still implicitly depend on the neuron variance. The simplest such diagram is\footnote{Note that since we are working on the lattice, we have $\int\!\mathrm{d}x\to a\sum_\ell$. We are also suppressing the external neuron indices $j,k$, it being understood that the total diagram is proportional to $\delta^{jk}$.} 
\begin{equation}
	\begin{aligned}
		\raisebox{-0.4\height}{\includegraphics[width=0.2\textwidth]{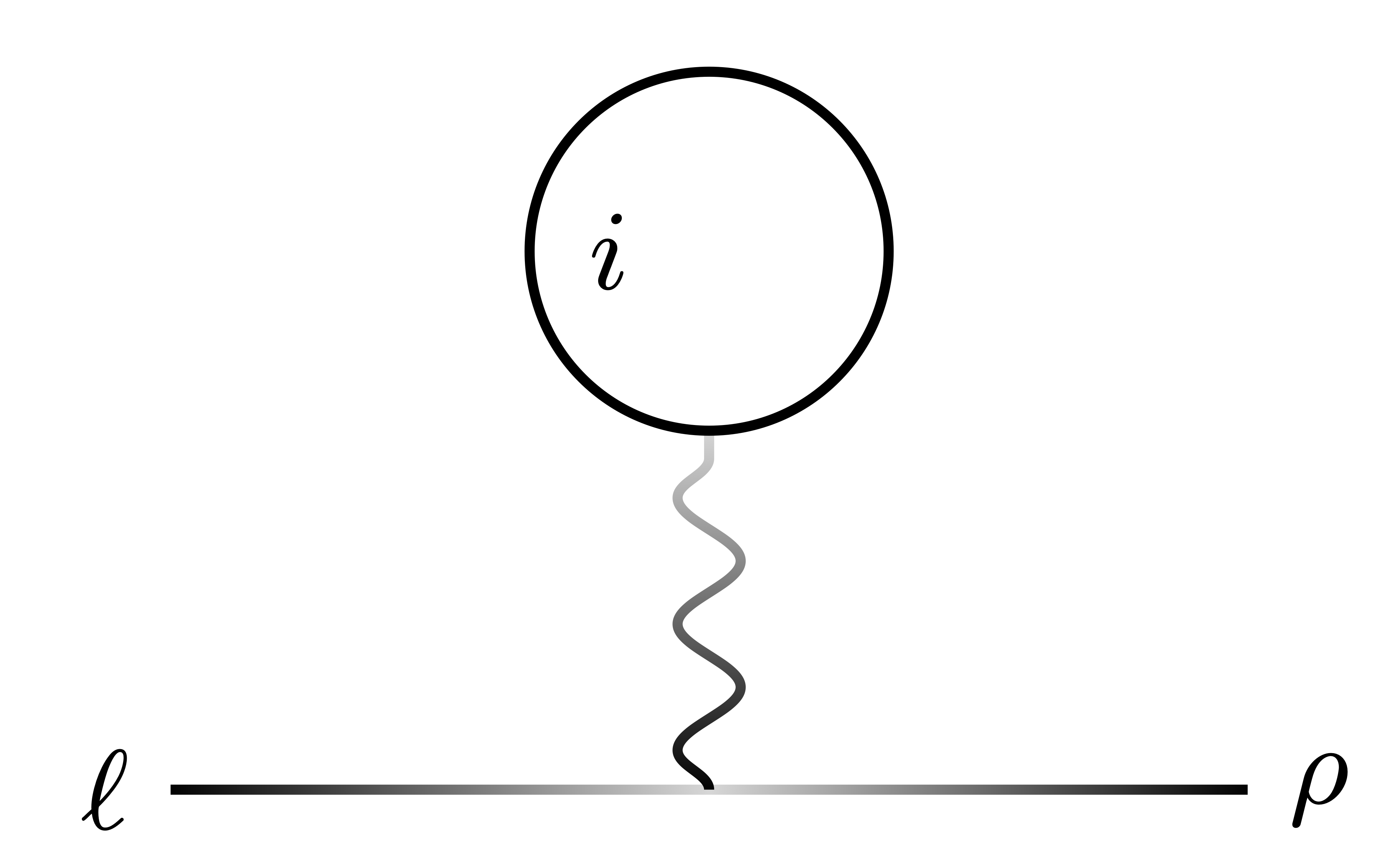}}&=N\left(-\frac{\gamma_2}{2}\frac{\sigma_w^2}{N}\right) a^2\sum_{u_1,u_2,u_3}G_{\ell u_1}^{hz}G_{u_1\rho}^{zh}G_{u_1u_2}^A\delta_{u_3,u_2-1}G_{u_3u_3}^{hh}\\
												 &=\frac{\gamma_2}{2}a^2\sigma_w^2\sum_{u_1,u_2,u_3}\delta_{\ell u_1}\delta_{u_1\rho}\delta_{u_1u_2}\delta_{u_3,u_2-1}\delta_{u_3u_3}\Delta_{u_3}\\
												  &=\frac{\gamma_2}{2}a^2\sigma_w^2\Delta_{\ell-1}\delta_{\ell\rho}~,
	\end{aligned}
	\label{eq:cac1}
\end{equation}
where the prefactor $N$ arises from the some over the contracted neuron index in the loop, and we have employed the obvious shorthand notation $G_{\ell\rho}^{hz}$ for the propagator (i.e., Green function) from $h_\ell$ to $z_\rho$, and similarly for the others. We have suppressed the labels on the internal vertices $u_i$ to avoid clutter, though it should be clear from context that they are labelled from bottom to top. We note that due to the aforementioned bilocality of the $Ah^n$ vertex, this diagram induces an effective dependence of the propagator at layer $\ell$ on that in the previous layer via the loop. 

Of course, the number of diagrams explodes as we increase the number of loops. At two loops, we have
\begin{equation}
	\raisebox{-0.4\height}{\includegraphics[width=0.2\textwidth]{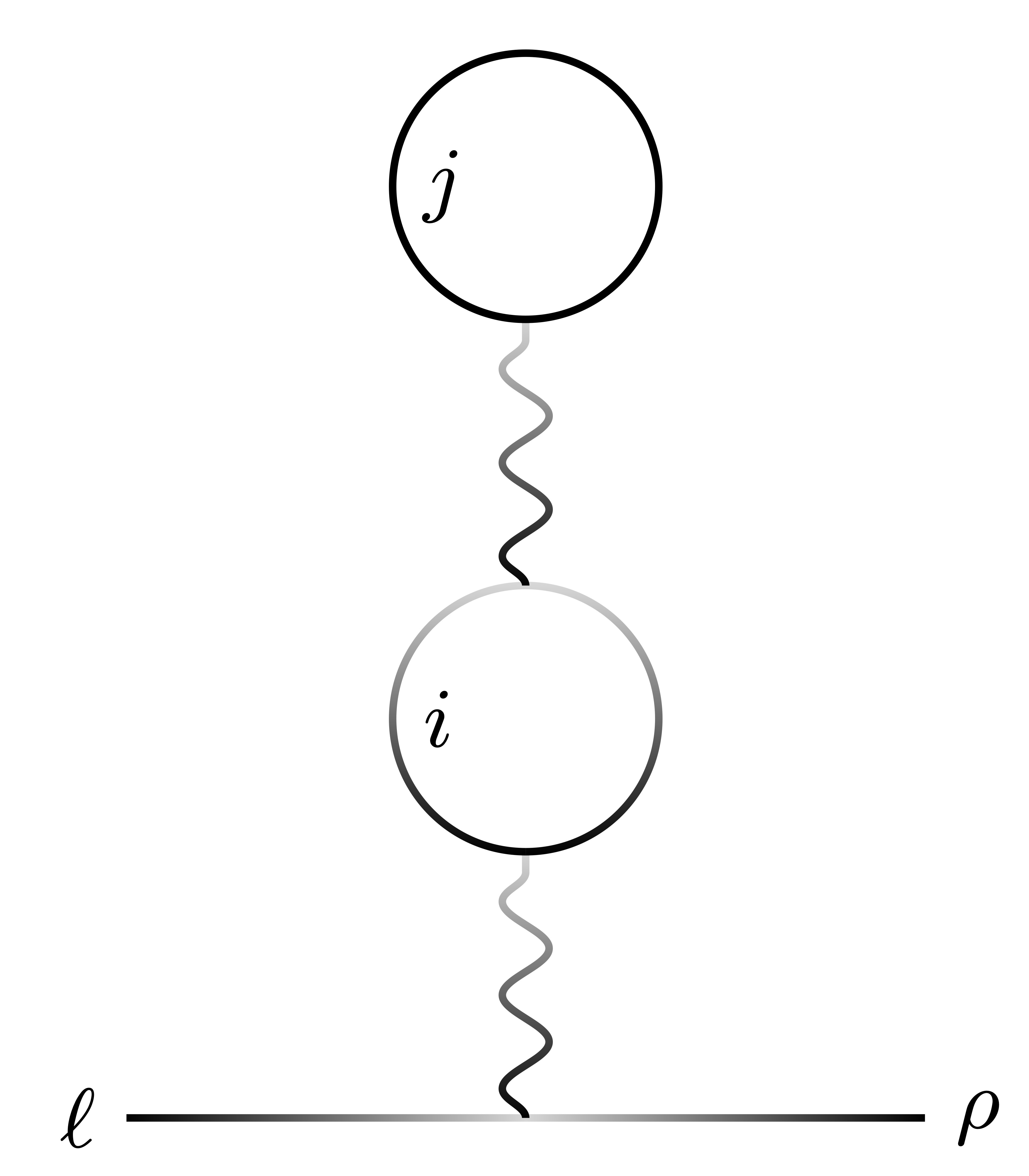}}
	\quad+\qquad
	\raisebox{-0.4\height}{\includegraphics[width=0.2\textwidth]{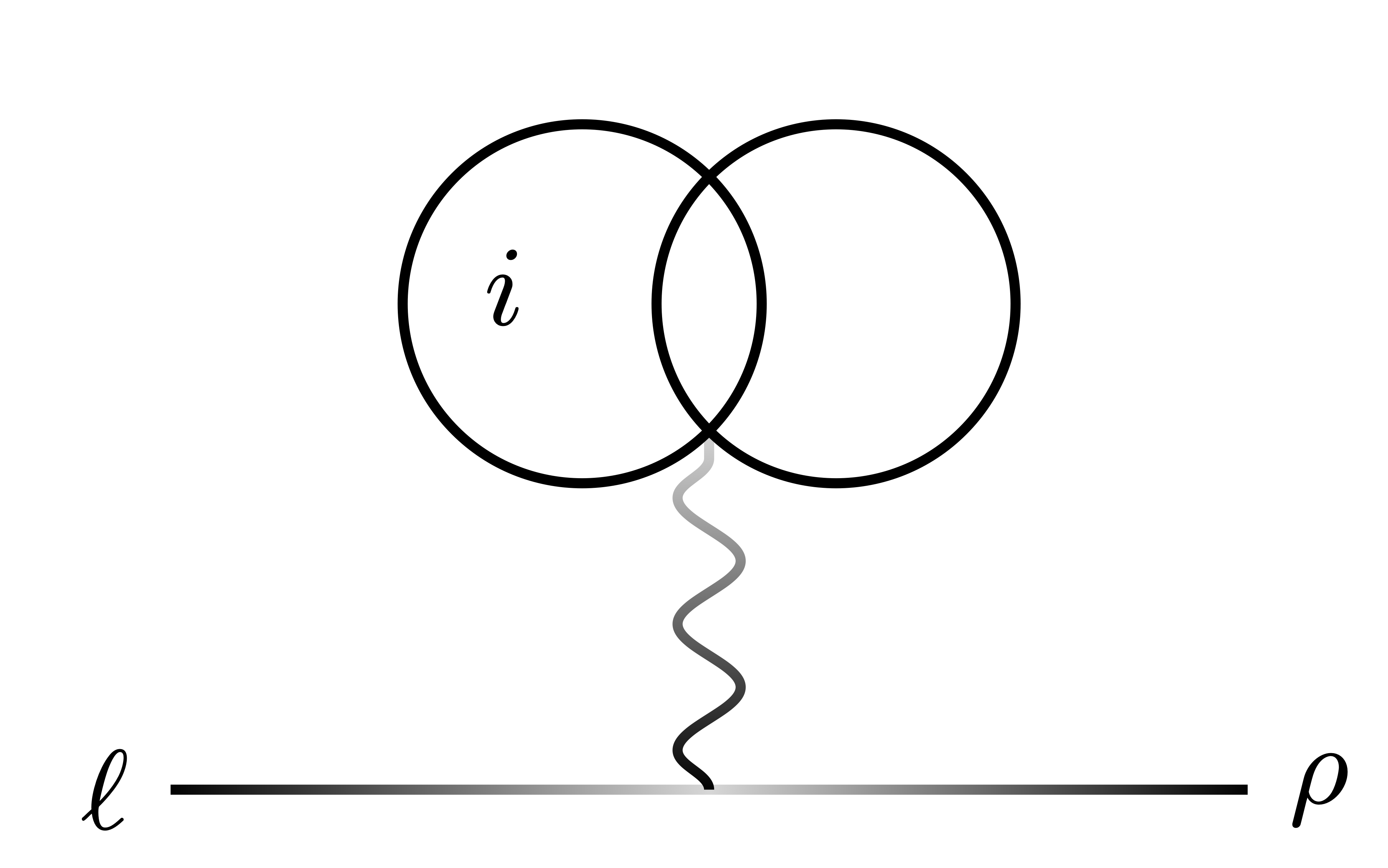}}~,
	\label{eq:n2}
\end{equation}
where it should be clear from context that there is no vertex where the $hh$ propagators happen to overlap, since such a vertex does not exist in the theory. Furthermore, since all neurons $h$ at the upper vertex in the second diagram share the same index, there is only a single factor of $N$ from both loops, so that the diagram remains $O(1)$. Note however that these diagrams come with symmetry factors associated with the possible contractions of internal legs: $1/2$ for the first (from the 2 possible ways to connect the two vertices comprising the lower loop) and $1/3$ for the second (due to the 3 ways to contract the four $h$ legs in the upper $Ah^4$ vertex). Additionally, the first diagram involves a dependence on $\Delta_{\ell-2}$, while the second diagram depends on $\Delta_{\ell-1}^2$. Thus we can already see that while all cactus diagrams are $O(1)$, higher loops correspond to an increasing degree of effective nonlocality.

At three loops, we begin seeing so-called branching cactii, where a stem can grow from either of the loops nearer the base:
\begin{equation}
	\raisebox{-0.5\height}{\includegraphics[width=0.2\textwidth]{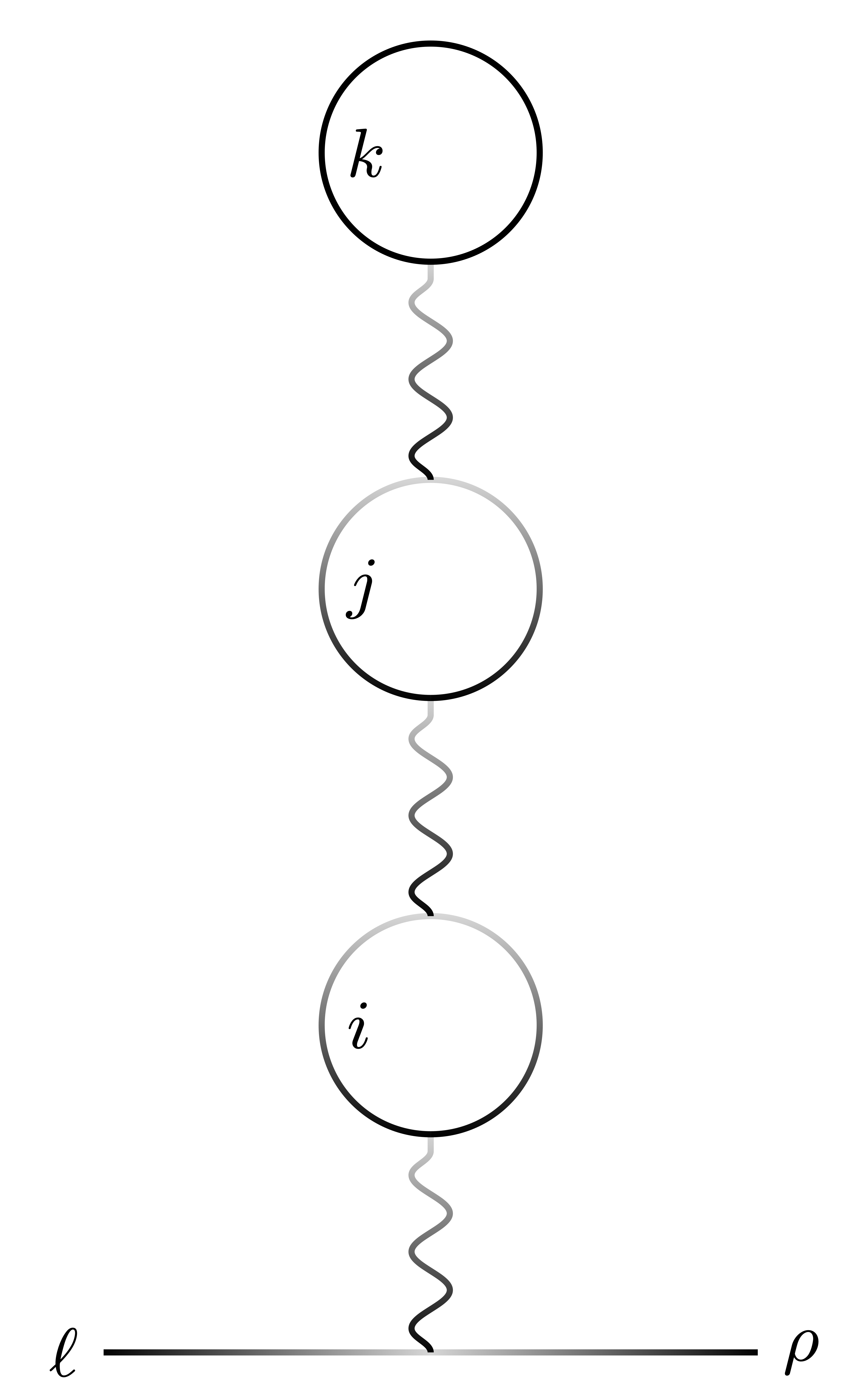}}
	+
	\raisebox{-0.4\height}{\includegraphics[width=0.2\textwidth]{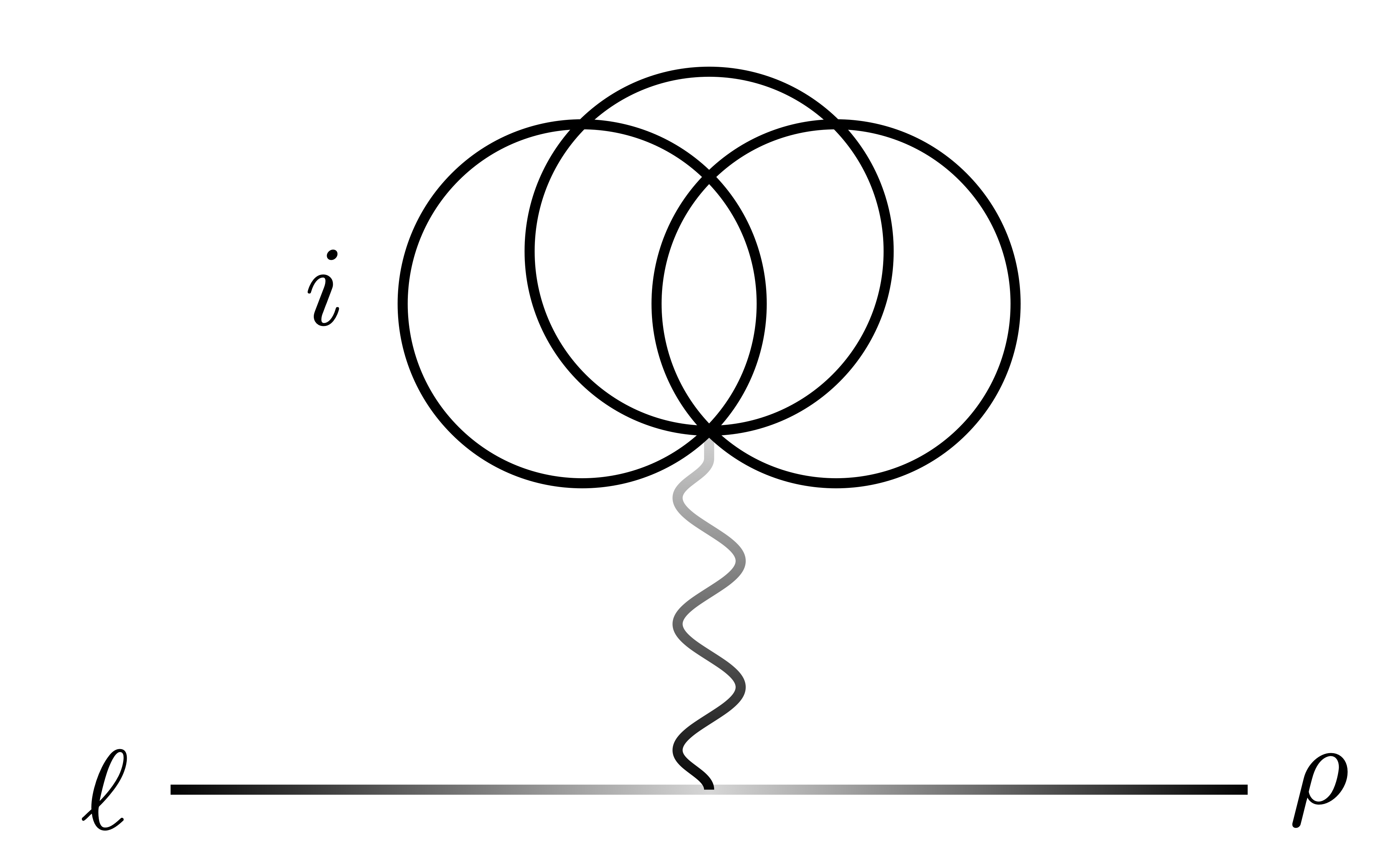}}
	+
	\raisebox{-0.4\height}{\includegraphics[width=0.2\textwidth]{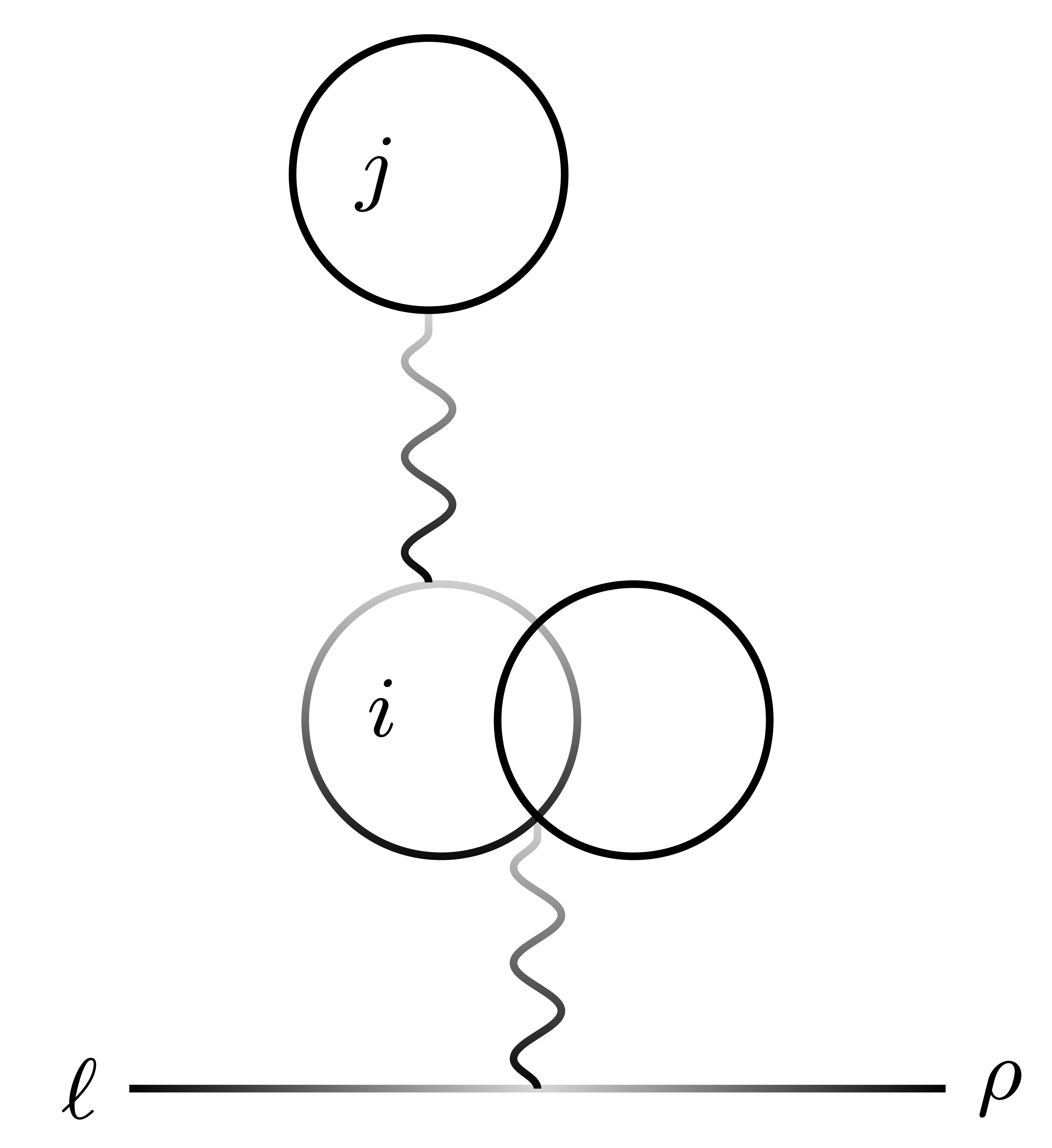}}
	+
	\raisebox{-0.4\height}{\includegraphics[width=0.2\textwidth]{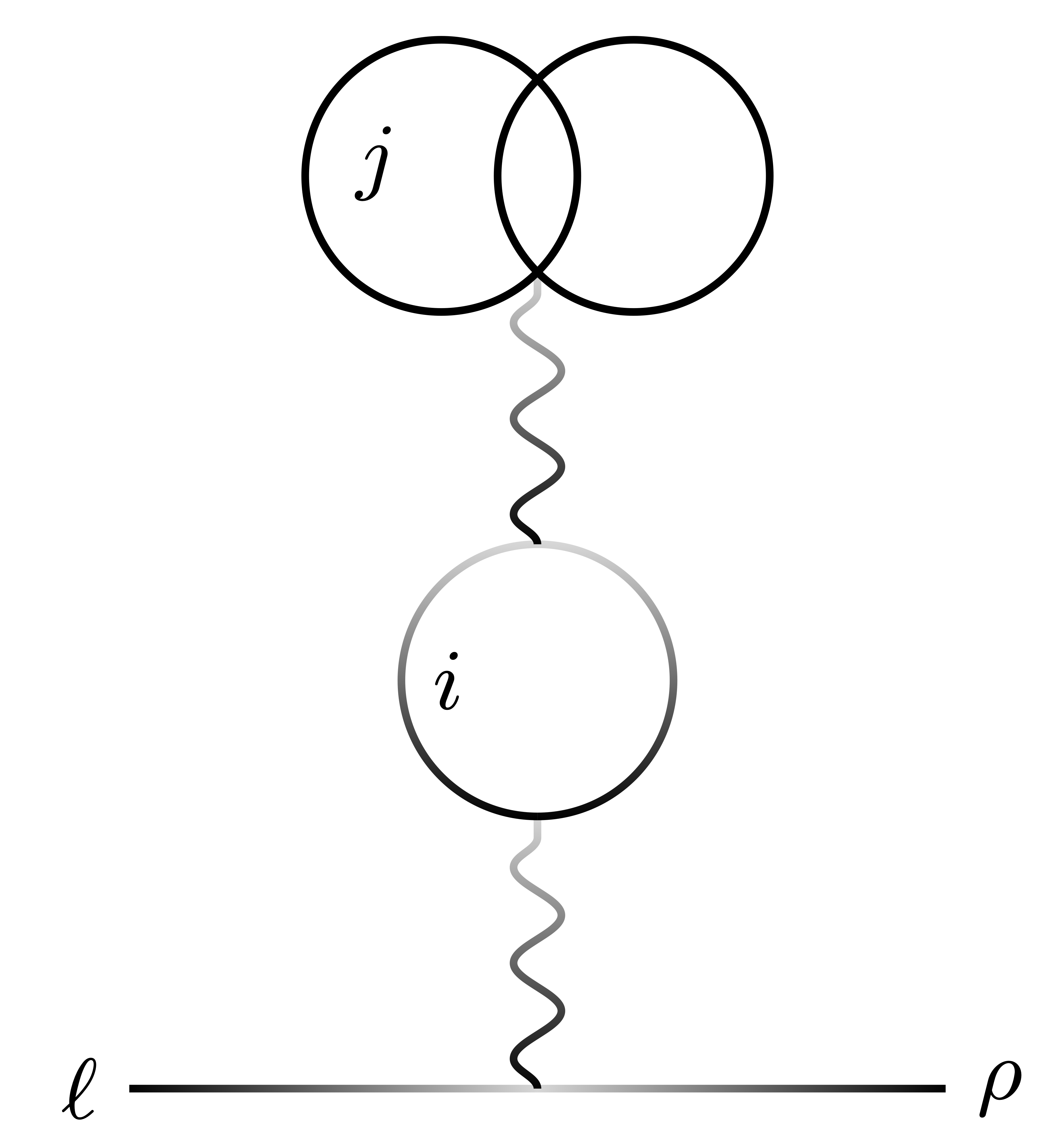}}~.
	\label{eq:n3}
\end{equation}
The associated symmetry factors are $1/2^2=1/4$ for the first diagram and $1/(6\!-\!1)!!\!=\!1/15$ for the second, for the same reasons as above. For the third, we obtain $1/12$ (which one can understand as 15 possible contractions of 6 internal legs minus 3 which yield disconnected graphs), while the last has $1/2\cdot 1/3=1/6$. Meanwhile the effective non-locality is of the form, respectively, $\Delta_{\ell-3}$ for the first, $\Delta_{\ell-1}^3$ for the second, $\Delta_{\ell-1}\Delta_{\ell-2}$ for the third, and $\Delta_{\ell-2}^2$ for the last.

To manage this unwieldy plethora, we employ a similar recursive formalism as in \cite{Grosvenor:2021eol}. Let us denote the exact propagator at layer $\ell$ by $X_\ell$, as distinct from the corresponding bare propagator $\Delta_\ell$. The contribution from all $O(1)$ diagrams above can then be expressed diagrammatically as
\begin{equation}
	\begin{aligned}
		\raisebox{-0.45\height}{\includegraphics[width=0.135\textwidth]{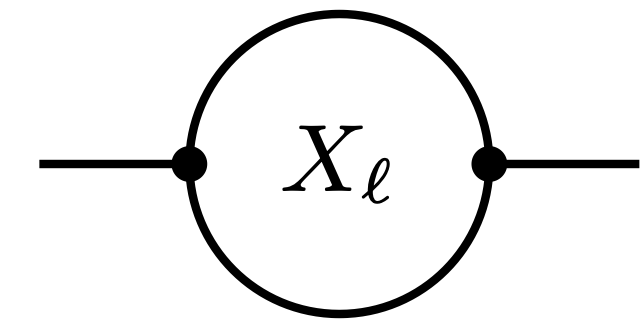}}
		\;\;=&\;\;
		\raisebox{-0.45\height}{\includegraphics[width=0.135\textwidth]{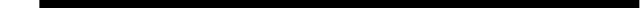}}
		\;\;+\;\raisebox{-0.45\height}{\includegraphics[width=0.18\textwidth]{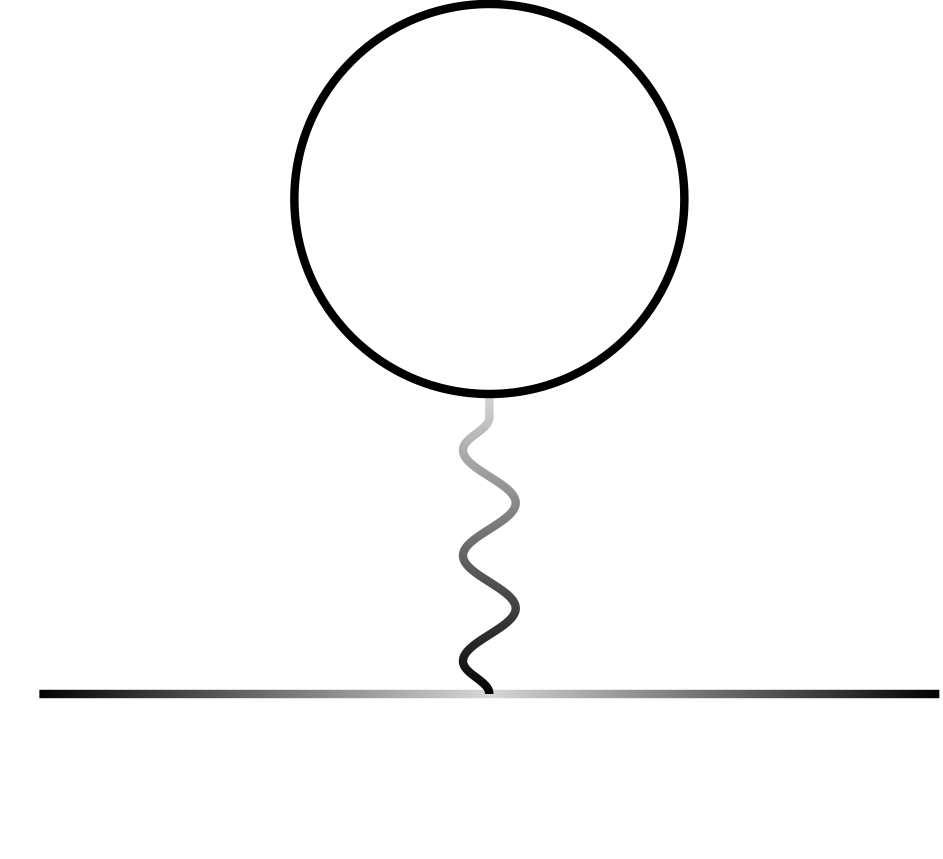}}
		\;\;+\;\raisebox{-0.45\height}{\includegraphics[width=0.18\textwidth]{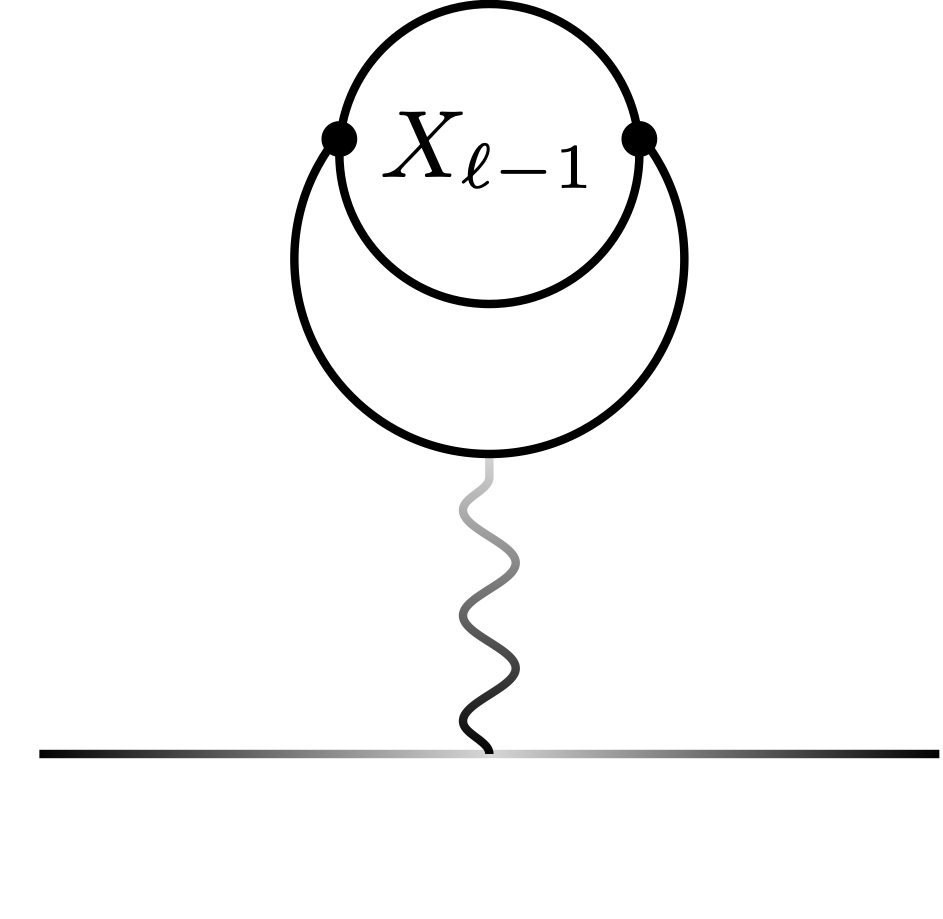}}\\
		 &+\;\raisebox{-0.4\height}{\includegraphics[width=0.18\textwidth]{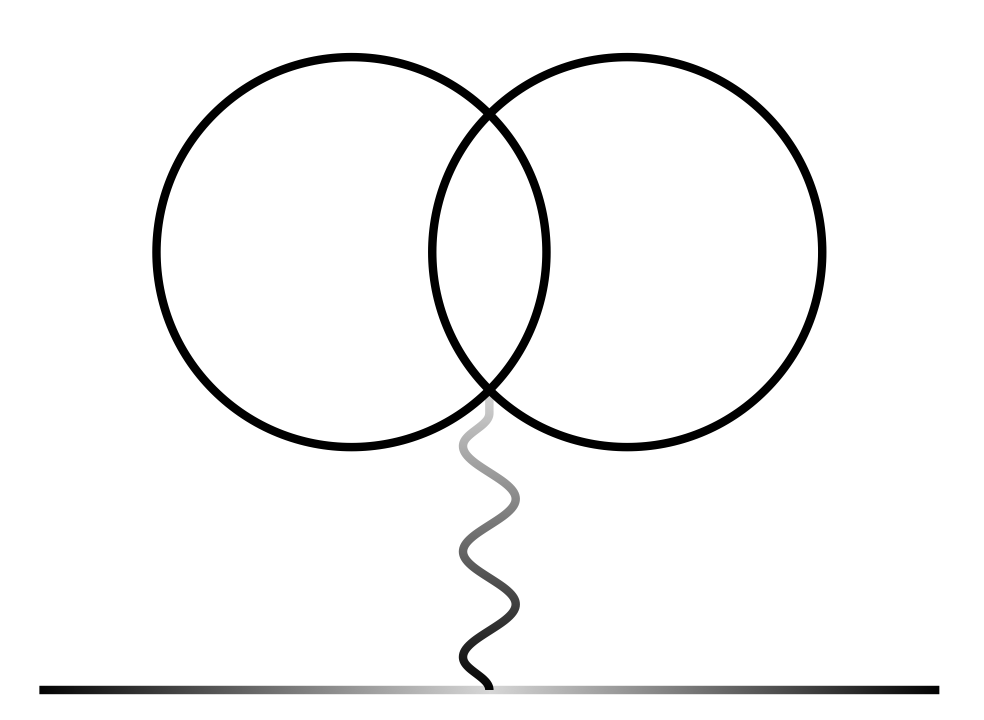}}
		 \quad+\quad\raisebox{-0.4\height}{\includegraphics[width=0.18\textwidth]{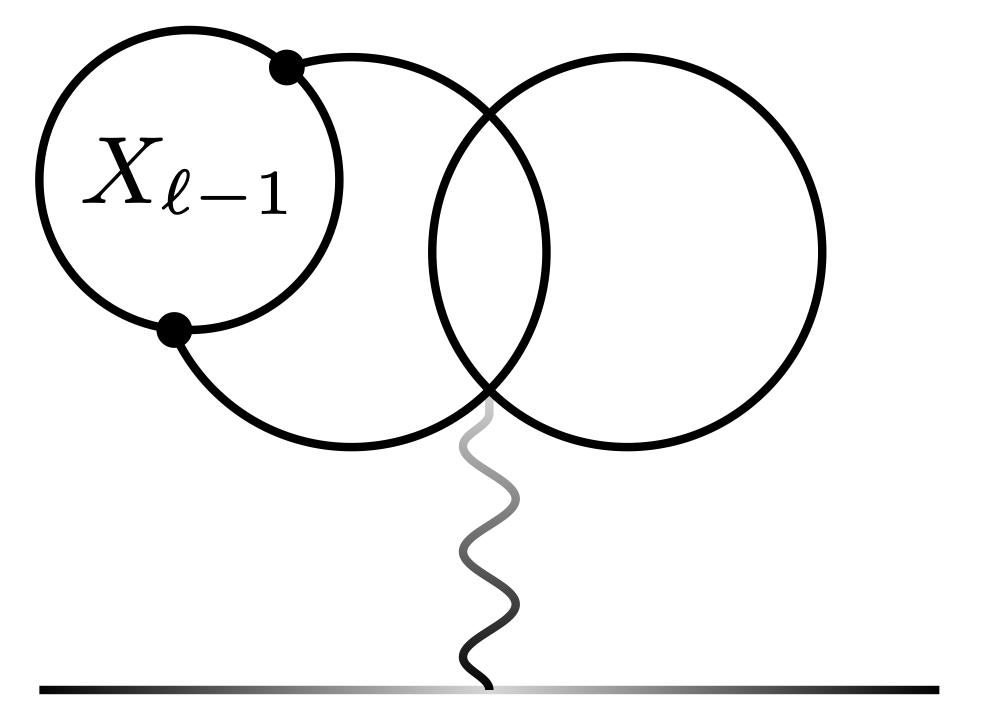}}
		 \quad+\quad\raisebox{-0.4\height}{\includegraphics[width=0.18\textwidth]{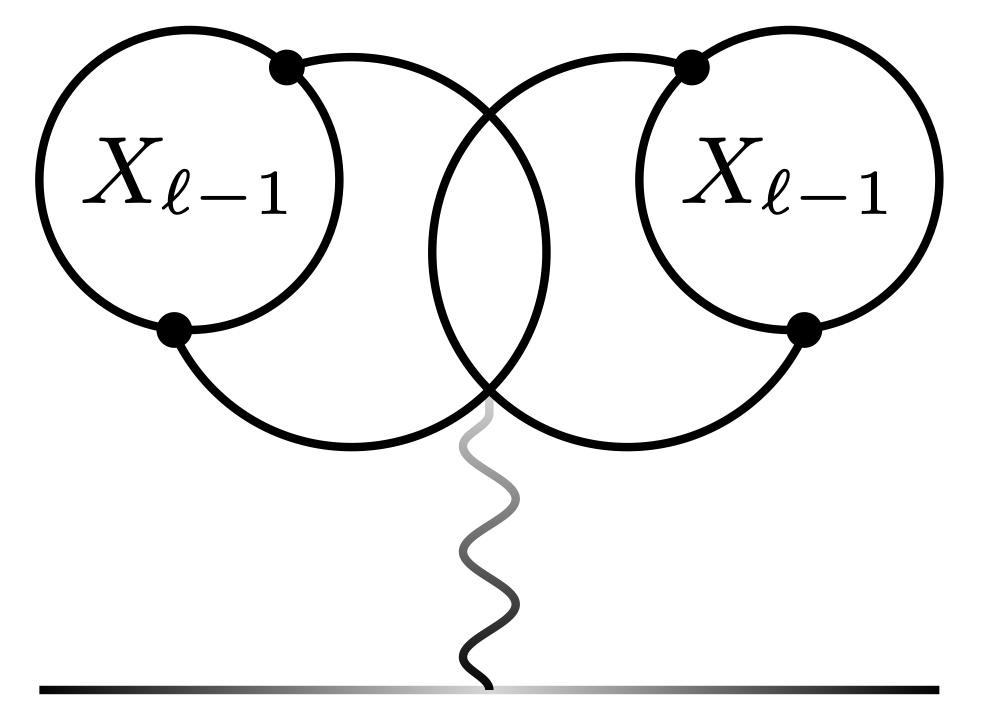}}
		 \;+\;\;\cdots~,
	\end{aligned}
	\label{eq:fun}
\end{equation}
in which $O(1)$ loop contributions are recursively substituted to yield the various cactus diagrams above. For example, substituting the second diagram on the right-hand side for $X_{\ell-1}$ in the third generates the $n\!=\!2$ cactus (the first diagram in \eqref{eq:n2}), while recursively substituting the second diagram into itself allows one to generate the entire tower of such cactii. Note that each additional vertical loop has the corresponding layer index reduced by one, so that the height of the towers grows for deeper networks.

Importantly, the reason we have split the diagrams in this way -- with each loop containing either a bare propagator, or a propagator dressed with what we shall refer to as a \emph{recursion node} (or simply a \emph{node}) $X_{\ell-1}$ -- is that it is necessary in order to reproduce the correct symmetry factors as exemplified above. For example, the second diagram on the right-hand side, with 1 loop and 0 nodes, has symmetry factor 1, while the third, with 1 loop and 1 node, has symmetry factor $1/2$ (obtained by treating the node as a vertex with two legs, thus giving 2 possible contractions with the two $h$ legs from the vertex at the bottom of the loop). In general, a $k\!\geq\!1$ loop diagram with $0\!\leq\!n\!\leq\!k$ nodes has the symmetry factor
\begin{equation}
	\frac{(2k-2n)!}{(2k)!(2k-2n-1)!!}=2^{k-n}\frac{(k-n)!}{(2k)!}\in(0,1]~,
	\label{eq:symfacnl}
\end{equation}
which attains its maximum value of $1$ only for the case $k\!=\!1$ and $n\!=\!0$, and approaches $0$ as $k\to\infty$.

It is then straightforward to obtain a formal expression for this recursive series by observing that the sum of loop diagrams on the right-hand side of \eqref{eq:fun} is formally similar to the following sum of so-called \emph{flower diagrams} (characterized by only a single $A\frak{A}$ propagator),
\begin{equation}
	\raisebox{-0.4\height}{\includegraphics[width=0.2\textwidth]{tower1.png}}
	\;+\;
	\raisebox{-0.4\height}{\includegraphics[width=0.2\textwidth]{flower2.png}}
	\;+\;
	\raisebox{-0.4\height}{\includegraphics[width=0.2\textwidth]{flower3.png}}
	\;+\;
	\cdots
	\label{eq:flowers}
\end{equation}
but with extra diagrams in which one or more of the internal loops is dressed with a recursion node. From \eqref{eq:cac1}, we see that an $k$-loop flower diagram gives the contribution 
\begin{equation}
	\textrm{flower}_\ell^k\;=\;\frac{1}{(2k-1)!!}\left(\frac{\gamma_{2k}}{2}a^2\sigma_w^2\right)\Delta_{\ell-1}^k
\end{equation}
where the double factorial is the symmetry factor arising from possible $hh$ contractions (consistent with \eqref{eq:symfacnl} with $n\!=\!0$). The expression for a general $n$-node flower diagram -- i.e., those appearing on the right-hand side of the recursion relation \eqref{eq:fun} -- is then obtained by replacing $n$ of the bare $\Delta_{\ell-1}$ propagators with nodes $X_{\ell-1}$, and replacing the symmetry factor with the general formula \eqref{eq:symfacnl}:
\begin{equation}
	\textrm{flower}_{\ell,n}^k\;=\;\left(\frac{\gamma_{2k}}{2}a^2\sigma_w^2\right)\frac{(k-n)!}{(2k)!}(2\Delta_{\ell-1})^{k-n}X_{\ell-1}^n~,
\end{equation}
which reduces to the previous formula when $n\!=\!0$. Thus, the recursive expression \eqref{eq:fun} for the exact 2-pt function, including all $O(1)$ corrections from arbitrarily-many loops, is
\begin{equation}
	X_{\ell}=\Delta_\ell+\frac{a^2\sigma_w^2}{2}\sum_{k=1}^\infty\frac{\gamma_{2k}}{(2k)!}(2\Delta_{\ell-1})^k\sum_{n=0}^k(k-n)!\left(\frac{X_{\ell-1}}{2\Delta_{\ell-1}}\right)^n~.
	\label{eq:Xfull}
\end{equation}
As implied below \eqref{eq:fun}, at layer $\ell\!=\!1$, only the non-recursive flower diagrams (those appearing in \eqref{eq:flowers}, with no recursion nodes) appear; i.e., we have the initial condition $X_0=\Delta_0$.

Unfortunately, despite some effort, we have not managed to obtain a closed-form expression for this relation. We have however obtained a reorganization of the infinite series in appendix \ref{sec:valiant}, in which the sum over loops (which, as explained therein, is not equivalent to a na\"ive sum over $k$ due to the recursion structure) has been performed in exchange for a sum over an expansion of the coefficients at each vertex. In the course of doing so, we prove the convergence of \eqref{eq:Xfull}, and reveal an interesting pole structure arising from the zeta function expansion of the coefficients. We hope to explore this numerically in the near future.

Before proceeding to higher-point correlators in the next section, let us briefly comment on the subleading $O(1/N)$ corrections to the bare propagator. The Feynman rules for our lattice theory -- in particular the rule for the $A_\ell(h_{\ell-1}^i)^n$ vertex, which increments the layer index -- imposes a significant restriction on the space of possible diagrams, which rules-out internal propagators that would connect vertices with incompatible layer indices. Nonetheless, we can construct an infinite zoo of $O(1/N)$ diagrams by joining stems of the cactii above subject to this constraint, which amounts to having two $\frak{A}A$ stems joined at an internal $hh$ loop. For example, the following two diagrams both contribute at $O(1/N)$:
\begin{equation}
	\begin{aligned}
		\raisebox{-0.4\height}{\includegraphics[width=0.2\textwidth]{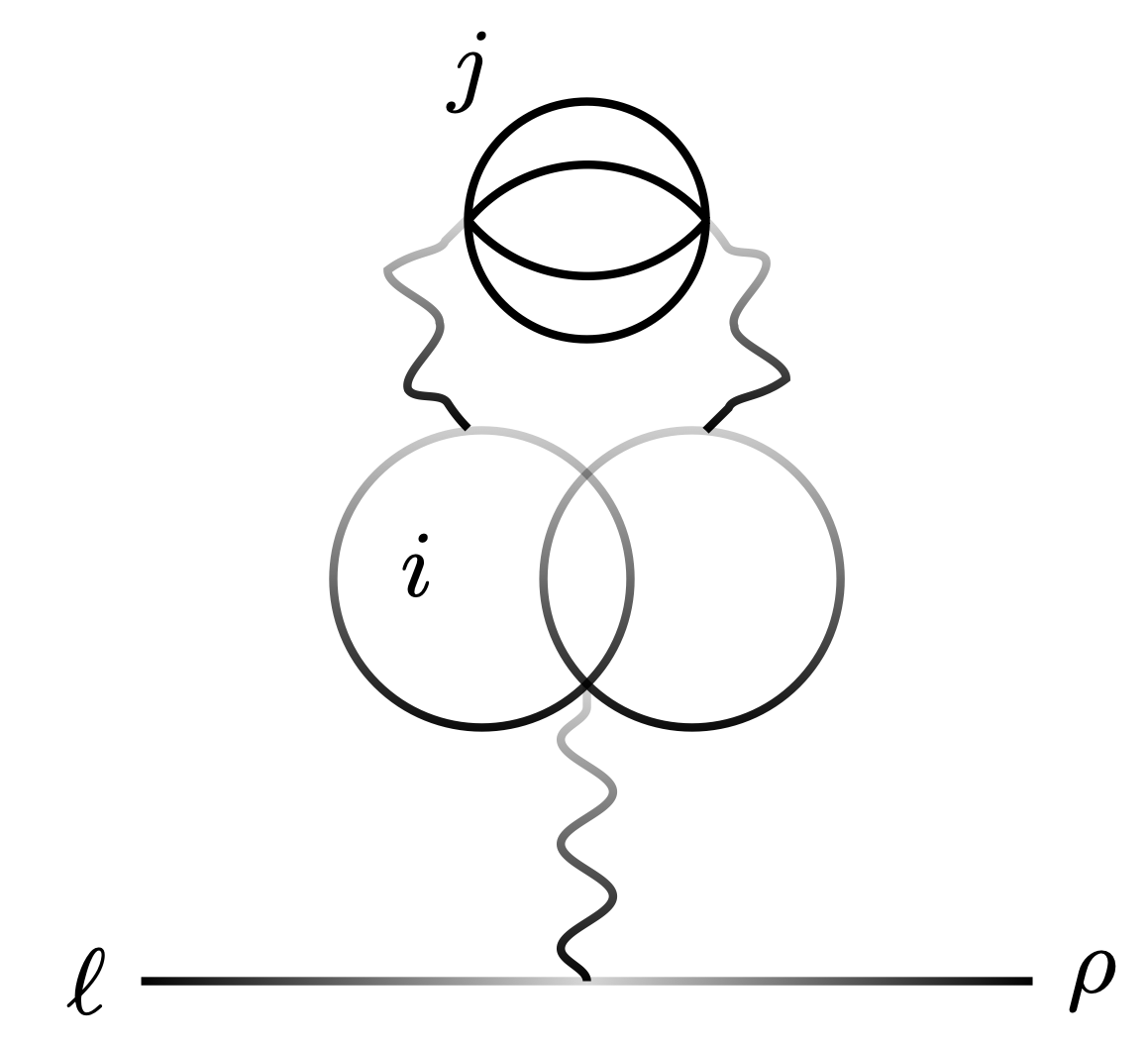}}\quad&\sim\quad\gamma_4^3\,\frac{\sigma_w^6}{N}\Delta_{\ell-2}^4\delta_{\ell\rho}~,\\
		\vspace{1em}
		\raisebox{-0.4\height}{\includegraphics[width=0.2\textwidth]{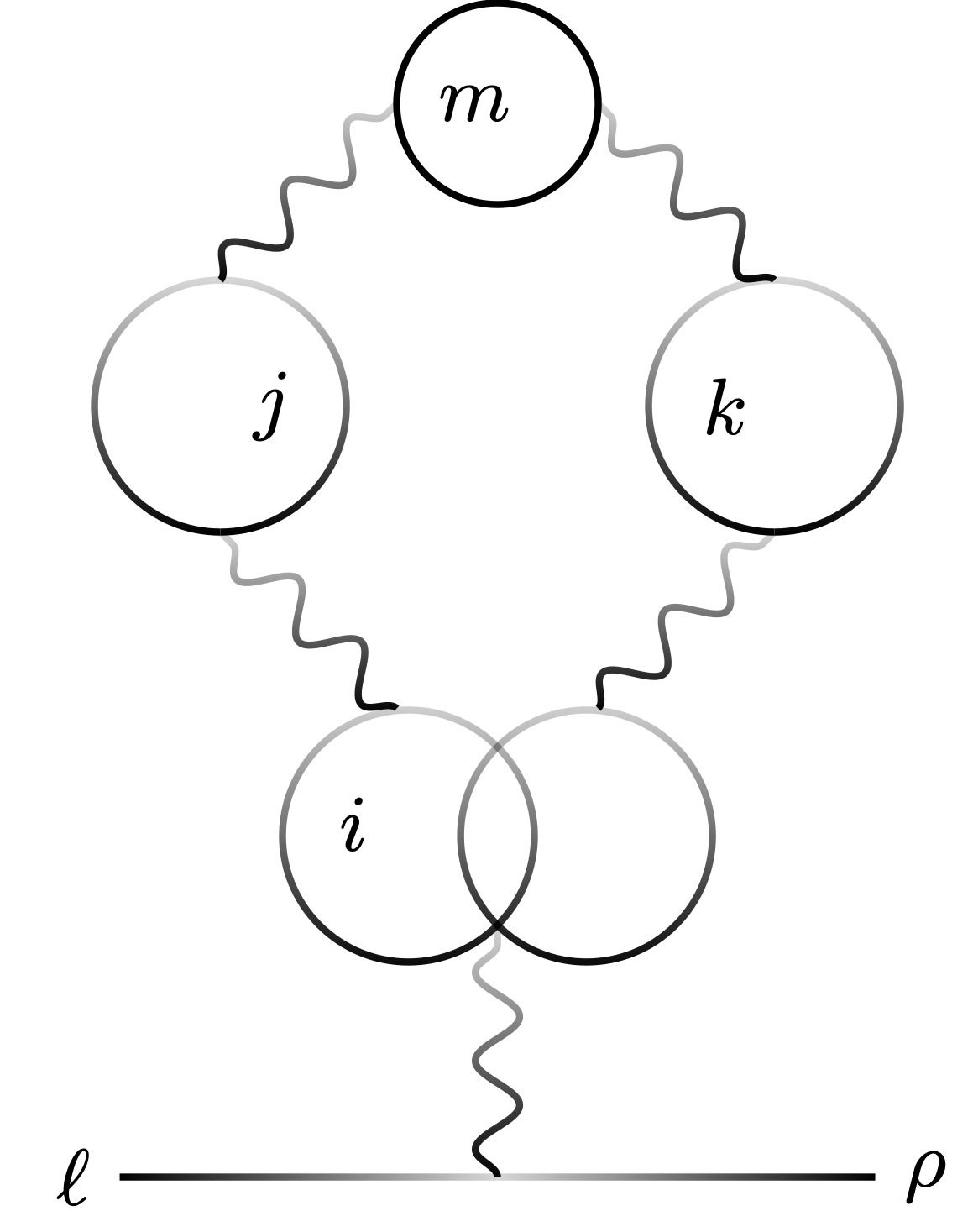}}\quad&\sim\quad\gamma_2^4\gamma_4\,\frac{\sigma_w^{10}}{N}\Delta_{\ell-3}^2\delta_{\ell\rho}~,
	\end{aligned}
\end{equation}
where we have suppressed symmetry factors for compactness. Note that in all such diagrams, the finite-width corrections to the propagator at layer $\ell$ are mediated via the propagators at previous layers, reflecting the causal nature of information flow in the network at initialization. Additionally, we observe that joining three or more stems in this manner would be further suppressed in $1/N$, since this would increase the number of $\frak{A}A$ propagators (which carry factors of $1/N$) relative to the number of internal neuron loops (which carry factors of $N$).

\section{Neuron scattering amplitudes}\label{sec:scatter}

As in the familiar gauge theories studied in physics, only gauge-invariant quantities are physical observables. In the present case, this implies that when considering higher-point correlation functions, the only non-vanishing observables are permutation-invariant combinations of neurons, such as
\begin{equation}
	\left<h_\ell^\mt h_\ell h_\rho^\mt h_\rho\right>~,
	\qquad
	\left<h_\ell^\mt h_\ell h_\rho^\mt h_\rho h_\tau^\mt h_\tau\right>~,
\end{equation}
and so on. As expected from the large-$N$ thermalization-like behaviour of the theory, there are no scattering amplitudes at $O(1)$: all $n$-point functions with $n>2$ will be at most $O(1/N)$, so that only the 2-point correlators survive in the infinite-width limit. In this section, we qualitatively analyze the predicted form of these higher-point functions in the perturbative expansion in $1/N$.

Let us begin by examining the tree-level diagrams allowed by our lattice Feynman rules. Thanks to the simplicity thereof, there is only a single class of 1PI diagrams that contributes, corresponding to $\left<h_\ell^2h_{\ell-1}^n\right>$:
\begin{equation}
	\raisebox{-0.4\height}{\includegraphics[width=0.3\textwidth]{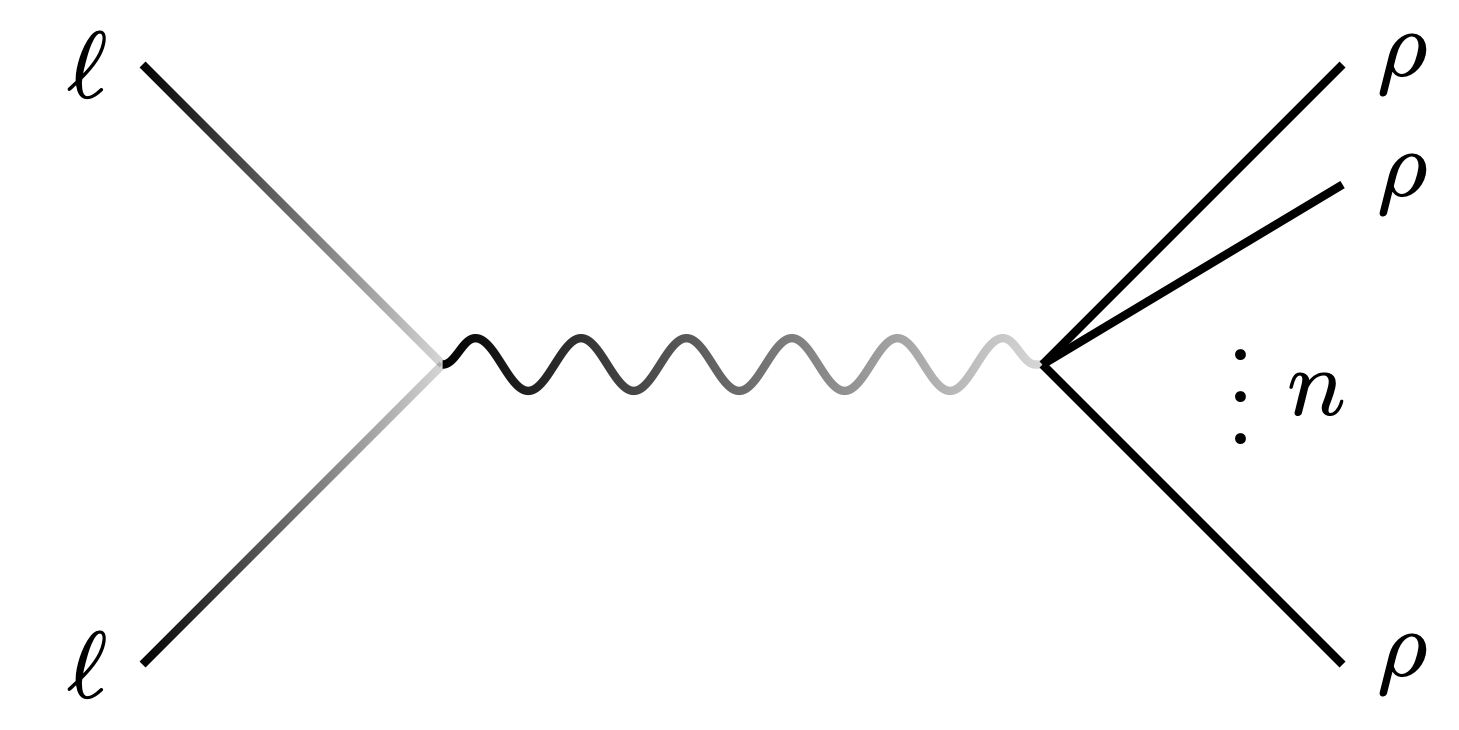}}
	\quad=\quad\frac{\gamma_n}{2}\frac{\sigma_w^2}{N}\delta_{\rho,\ell-1}~.
\end{equation}
Note that this diagram is $O(1/N)$, and that any attempt to construct a tree-level diagram that is $O(1/N^2)$ or higher would not be 1PI.\footnote{Since there are no $zz$ propagators, cutting an internal $hz$ propagator must result in an $hh$ leg and another $hz$ leg; similarly when cutting internal $zh$ propagators.} Curiously, there are no tree-level contributions to $\left<h_\ell^m\right>$ for $m>2$: as discussed in subsec. \ref{sec:rules}, the $A_\ell(h_{\ell-1})^n$ vertex involves a change of the layer index, such that the tree-level diagrams couple adjacent layers. 

For loop diagrams, let us proceed order-by-order in $1/N$. The contributions at $O(1/N)$ are obtained by essentially adding external legs to the top of the single-tower cactus diagrams considered in sec. \ref{sec:perturb}. These come in two classes: those joining internal $\frak{A}A$ propagators via a mixed ($hz$ and $zh$) loop, and those joining them via a single $hh$ loop:
\begin{equation}
	\begin{aligned}
	\raisebox{-0.4\height}{\includegraphics[width=0.4\textwidth]{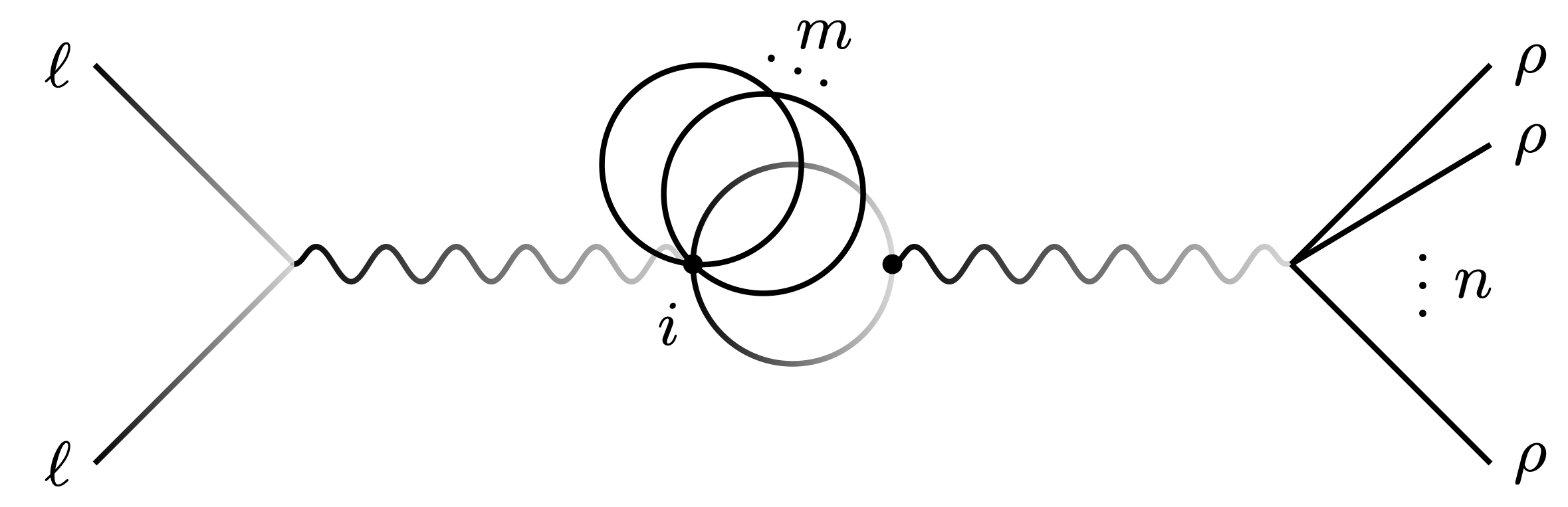}}\quad&=\quad\frac{\sigma_w^4}{N}\frac{\gamma_m\gamma_n}{4m(m-1)(m-3)!!}\Delta_{\ell-1}^{(m-2)/2}\delta_{\rho,\ell-2}~,\\
	\raisebox{-0.4\height}{\includegraphics[width=0.4\textwidth]{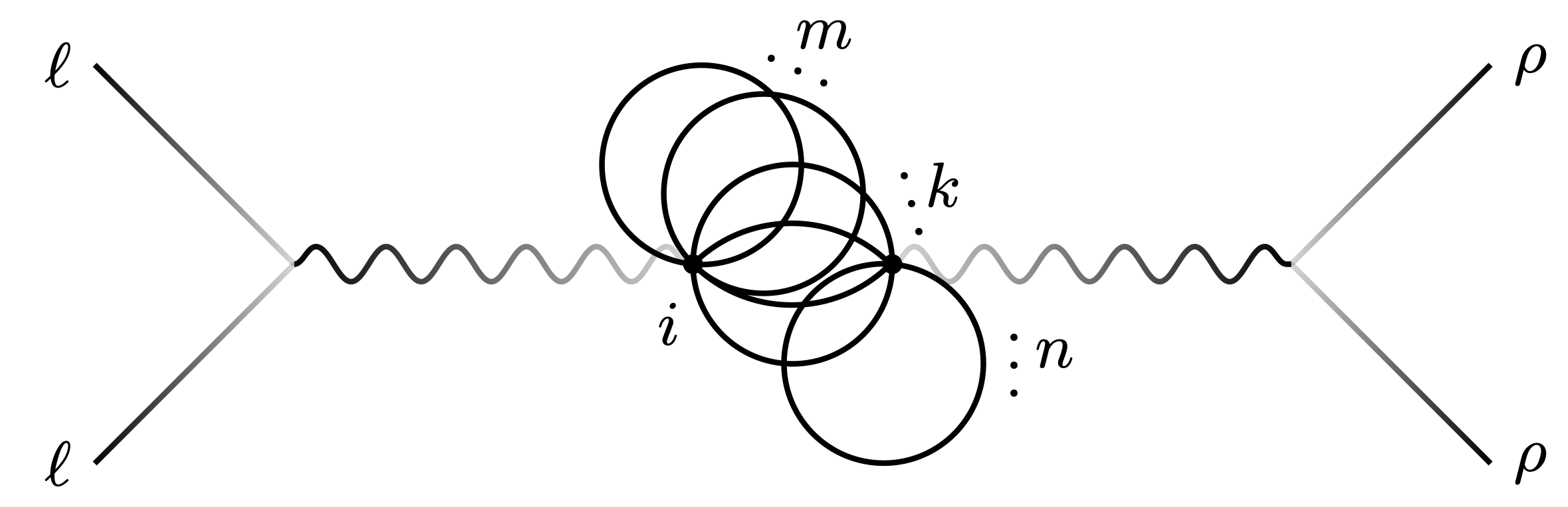}}\quad&=\quad\frac{\sigma_w^4}{N}\frac{\gamma_m\gamma_nk!}{4\lceil n,m\rceil!(m-k-1)!!(n-k-1)!!}\Delta_{\ell-1}^{(m+n)/2}\delta_{\rho,\ell}~.
	\end{aligned}
	\label{eq:scatter1}
\end{equation}
In these diagrams, $n,m$ are the numbers of legs attached to the corresponding vertex, while $k$ is the number of intra-vertex connections in the second diagram. Note that both of these diagrams contain only a single sum over the internal free neuron indexed by $i$. In the symmetry factor for the second diagram, $\lceil n,m\rceil$ denotes whichever of $n,m$ is larger, and arises from the possible contraction structure.\footnote{For example, if $m>n$, then there are $m!/k!$ possible contractions involving $k$ propagators between the two vertices, leaving $(m\!-\!k\!-\!1)!!$ pairs for the $m$-vertex and $(n\!-\!k\!-\!1)!!$ pairs for the $n$-vertex.} We have also added dots showing the vertices for clarity.

Qualitatively, there is an important difference between these types of diagrams, namely that the first couples two disparate layers (in this case, $\left<h_\ell^2h_{\ell-2}^n\right>$, though clearly $\left<h_\ell^nh_{\ell-2}^2\right>$ is also possible), while the second represents the first correction to a higher-point amplitude involving same-layer insertions, i.e., $\left<h_\ell^4\right>$. Note that in both cases, the scattering is mediated by previous layers via $\Delta_{\ell-1}$. Of course, we can add additional layers internally to both classes, but note that only a single $hh$ joint is possible:
\begin{equation}
	\begin{aligned}
		\raisebox{-0.4\height}{\includegraphics[width=0.8\textwidth]{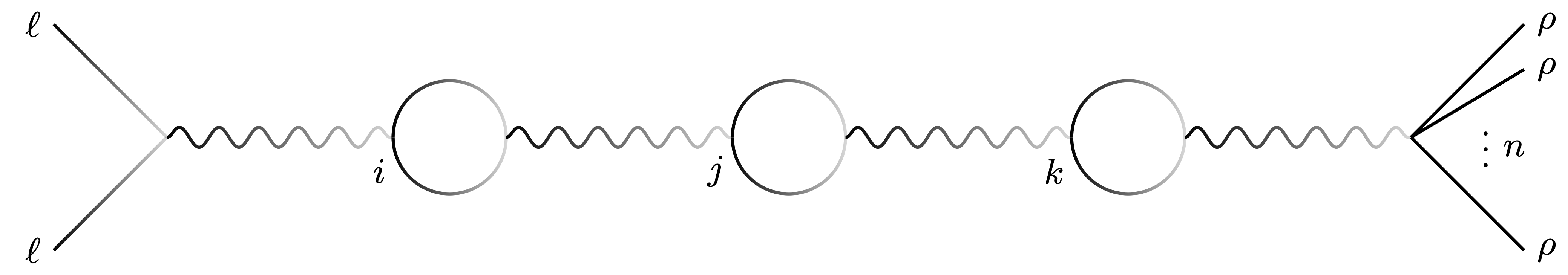}}\;&\sim\;\frac{\sigma_w^8}{N}\delta_{\rho,\ell-4}~,\\
		\raisebox{-0.4\height}{\includegraphics[width=0.8\textwidth]{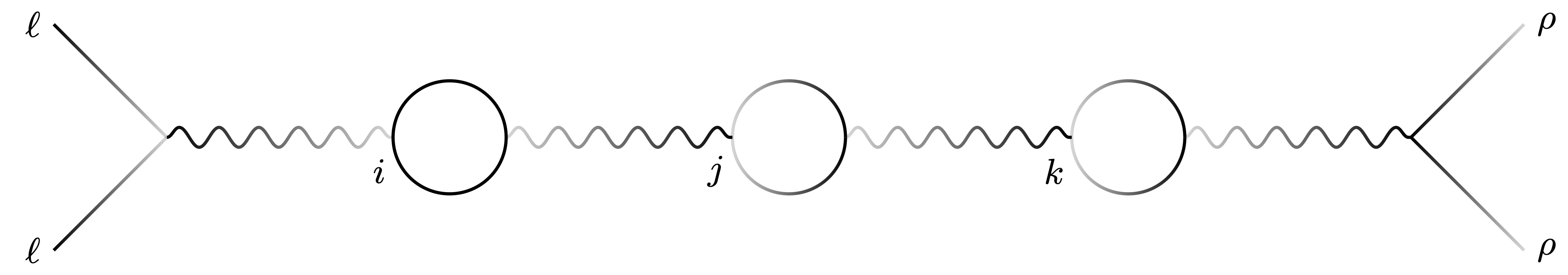}}\;&\sim\;\frac{\sigma_w^8}{N}\Delta_{\ell-1}^2\delta_{\rho,\ell+2}~,
	\end{aligned}
\end{equation}
where we have suppressed the capacity to place additional self-contractions on the loops as in \eqref{eq:scatter1} to avoid clutter. In all cases, we note that these diagrams contribute at $O(1/N)$, regardless of how far-separated are the external layers $\ell$ and $\rho$, since each additional distinct neuron loop (factor of $N$) also contains an additional $\frak{A}A$ propagator (factor of $1/N$). Additionally, including a single $hh$ loop in the internal chain of propagators is restricted to 4-point functions $\left<(h_\ell)^2(h_\rho)^2\right>$; in contrast, a chain can include arbitrarily-many mixed $hz$ and $zh$ loops, which contribute to $\left<(h_\ell)^2(h_\rho)^n\right>$.

At $O(1/N^2)$, the spectrum of possible diagrams becomes substantially more complicated, since -- as should by now be apparent -- one has the ability to attach (loop and stem chains leading to) external legs to any of the internal vertices appearing in the myriad of cactus and scattering diagrams above. Importantly, one sees correlations between more than two layers enter at this order, e.g., $\left<h_\ell^2h_{\ell+1}^2h_{\ell+2}^2\right>$:
\begin{equation}
	\raisebox{-0.2\height}{\includegraphics[width=0.8\textwidth]{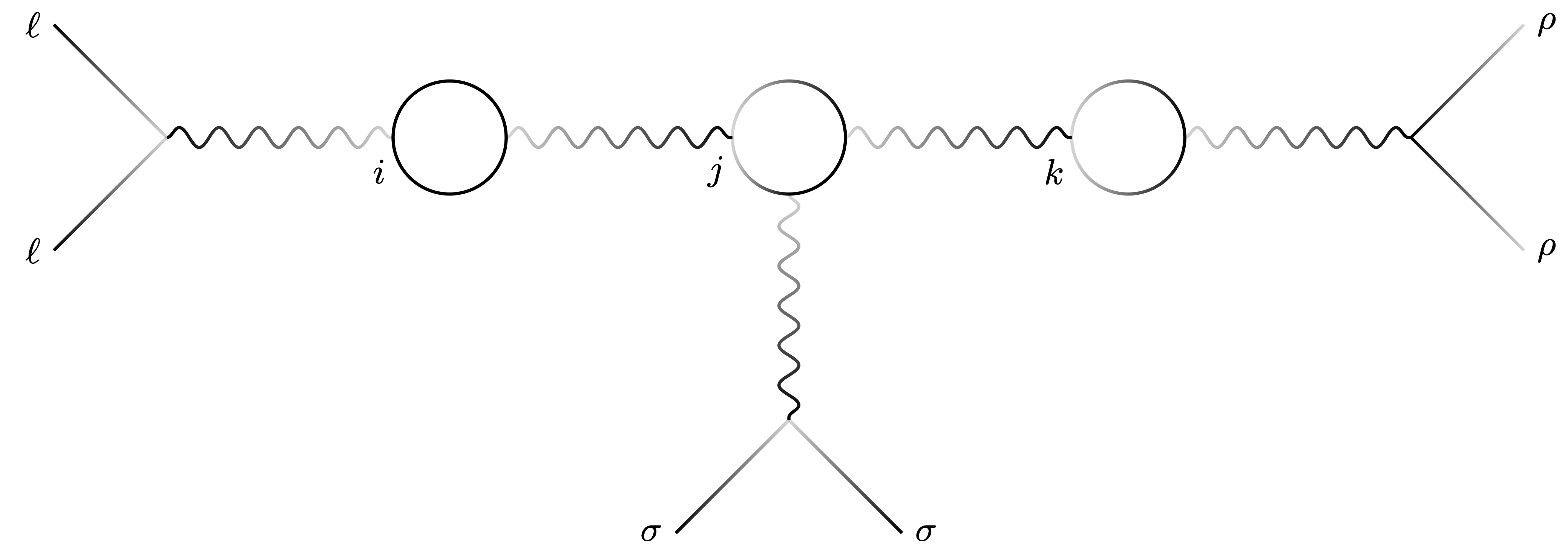}}
	\hspace{-4cm}\sim\frac{\sigma_w^{10}}{N^2}\Delta_{\ell-1}^2\Delta_\ell\delta_{\rho,\ell+2}\delta_{\sigma,\ell+1}~.
\end{equation}

\pagebreak 

\noindent Additionally at this order, it becomes possible to use internal $hh$ propagators to connect different loops, such as in the following contribution to $\left<(h_\ell)^4\right>$:
\begin{equation}
	\raisebox{-0.4\height}{\includegraphics[width=0.8\textwidth]{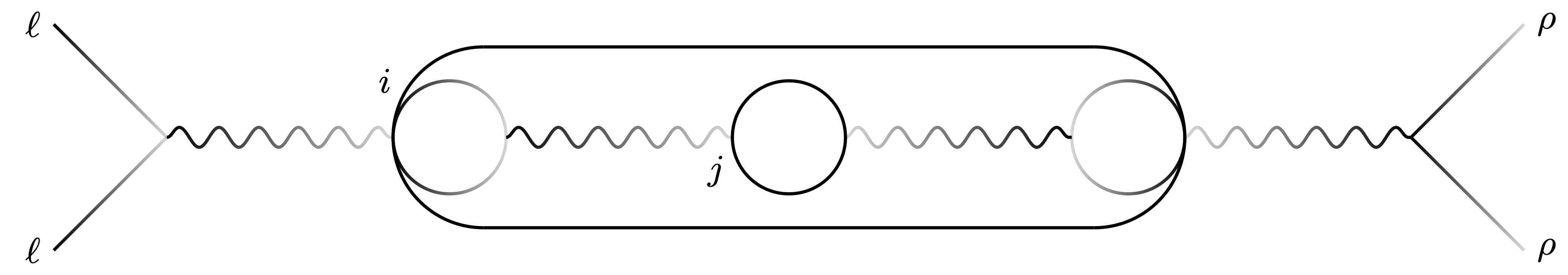}}
	\sim\;\frac{\sigma_w^8}{N^2}\Delta_{\ell-1}^2\Delta_{\ell-2}^2\delta_{\rho\ell}~.
\end{equation}

Collectively, the above suggests the following hierarchy of correlation functions: at $O(1)$, only the 2-point functions considered in the previous section survive. At $O(1/N)$, $n$-point functions of the form $\left<(h_\ell)^a(h_\rho)^b\right>$ with $a\!+\!b=n$ and $\rho\neq\ell$, which couple exactly two distinct layers, appear. At $O(1/N^2)$, one finally begins to see $n$-point functions involving more than two layers, in addition to new contributions to adjacent-layer $n$-point functions. Furthermore, note that the Feynman rules do not permit the construction of diagrams which violate the $S_N$ symmetry---in particular, recall that in the $A_\ell(h_{\ell-1})^n$ vertex, $n$ must be an even integer. 

\section{Discussion}\label{sec:discussion}

In this work, we have improved on the NN/QFT duality in \cite{Grosvenor:2021eol} to explicitly incorporate the layerwise permutation symmetry of fully-connected deep neural networks, resulting in a $(0\!+\!1)$-dimensional lattice gauge theory with discrete symmetry group $S_N$. Due to the low dimension of the theory -- in particular the lack of any square plaquette terms on the lattice -- there is no field strength term for the gauge fields, and the local symmetry manifests only in the covariantized derivative used in transcribing the structure of the network as an SDE. The Feynman rules are structurally similar to those in \cite{Grosvenor:2021eol}, but differ crucially due to both the local character of the weight matrices (in contrast to an RNN characterized by weight sharing) and the fact that we work on the lattice, which simplifies computations. In particular, expanding around the vev reveals the recursive nature of the diagrams, whereby correlations at layer $\ell$ may receive corrections from loops representing the influence of neurons in previous layers $\ell\!-\!1$, $\ell\!-\!2$, and so on, reflecting the causal nature of information propagation in networks at initialization (i.e., layer $\ell$ depends on layer $\ell\!-\!1$ but not on layer $\ell\!+\!1$).

In this framework, we have obtained recursive expressions for both the bare neuron-neuron propagator \eqref{eq:hhrec}, and the so-called quantum-corrected or exact propagator \eqref{eq:Xfull} to $O(1)$ in the perturbative expansion in $1/N$. As explained in \cite{Grosvenor:2021eol}, the $O(1)$ corrections represent statistical fluctuations in the ensemble of networks, and survive even in the infinite-width limit. A notable technical improvement relative to this previous work is that we have managed to keep infinitely-many terms in the expansion of the nonlinear activation function, treating this as a formal power series, which corresponds to retaining arbitrarily high $n$-point interactions. The resulting series contains an interesting pole structure analyzed in appendix \ref{sec:valiant}. However, we have confined our quantitative study of the propagator to this leading $O(1)$ effect. For the first genuinely finite-width correction, at $O(1/N)$, exemplary diagrams seem to indicate a breakdown at strong 't Hooft coupling $\sigma_w^2$, which may be consistent with earlier studies indicating a chaotic regime at large values of this parameter, see e.g. \cite{schoenholz2017deep,Bukva:2023ksv}.

We have also gone beyond previous works in examining higher-point correlation functions in the hidden layers of the network. While our investigations of these are preliminary, it suggests a potential approach to studying information propagation in the networks using tools from quantum field theory. In particular, we note that only gauge-invariant correlators are non-vanishing, and that the coupling between increasingly-many disparate layers appears to be suppressed in $1/N$: scattering between two layers $\ell$ and $\rho$ enters at $O(1/N)$, while that between three layers $\ell$, $\rho$, and $\sigma$ enters only at $O(1/N^2)$. It would be interesting to examine such neuron scattering amplitudes in more detail, but leave such an analysis to future work.

Speaking of which, let us remark on several potentially interesting directions. First, as mentioned above, the gauge fields are non-dynamical in the sense that they are not governed by any equation of motion but instead fluctuate randomly at each link. We expect that this is no longer true if one considers training, since stochastic gradient descent (SGD) gives rise to a kinetic term that evolves the weights based on the minimization of the external loss function. The extra temporal direction elevates the theory to a $(1\!+\!1)$-dimensional lattice, thereby circumventing the geometrical limitation to parallel transport in the present $(0\!+\!1)$-dimensional case. Of course, the ability to describe dynamical networks in this framework would also substantially expand the practical interest of this NN/QFT duality. We hope to explore this in the near future. 

Another direction of more theoretical interest concerns linear networks, since in that case the discrete permutation symmetry is elevated to a continuous rotation symmetry, with Lie group $\mathrm{SO}(N)$. In this case one is no longer constrained to work on the lattice, and can again work in the continuum limit. The resulting theory is still $(0\!+\!1)$-dimensional, but a preliminary analysis of the resulting Feynman rules reveals tantalizing connections with random matrix theory (RMT). In particular, in the course of our explorations, we found that for the continuum theory with $\mathrm{SO}(N)$ invariance, the $O(1)$ corrections to the propagator include so-called \emph{rainbow diagrams} formally identical to those which appear in computations of the 1PI self-energy in random matrix models \cite{PhysRevE.49.2588,ZEE1996726}; see also the discussion of the Wigner semicircle law in \cite{Zee:2003mt}. In light of the utility of RMT for machine learning \cite{couillet_liao_2022,RMTbeyond,bordelon2026disordereddynamicshighdimensions}, this could be worth a more detailed exploration in the future. We also note that, rather than limiting oneself to standard networks, one could consider modifying the activation function to preserve some desired symmetry as in \cite{iqbal2026spontaneoussymmetrybreakinggoldstone}, thereby widening the applicability well-beyond simple linear models. 

Furthermore, we have focused in the present work on fully-connected feedforward DNNs because they provide a simple (and by now standard) starting ground for such theoretical studies. It would be interesting to extend these methods to more complicated network architectures, such as CNNs or transformers. In the case of the former for example, one might expect a $(2\!+\!1)$-dimensional theory owing to the fact that unlike MLPs, CNNs retain spatial information within each layer. It would also be interesting to connect with other approaches based on statistical field theory, for example to shed further light on scaling laws \cite{bordelon2025theoryscalinglawsincontext,bordelon2024dynamicalmodelneuralscaling} in this framework.

Lastly of course, the theoretical findings in this work lend themselves to empirical validation on real-world networks. Implicit in this is the question of how to properly treat the lattice spacing $a$, which we have so far retained as a free parameter in the theory. One may also ask whether the covariantized ODE describing the structure of the network is indeed a faithful representation away from the continuum limit, or if this should instead be taken as a representation of a modified network with a ResNet-like connection controlled by the lattice spacing. We hope to report on these empirical explorations in a companion paper \cite{TBA} in the near future. 

\section*{Acknowledgments}

We thank Abbas Abbasli, David Berman, Blake Bordelon, Kevin Grosvenor, Koji Hashimoto, Moritz Helias, Boris Hanin, Nabil Iqbal, Joris Kieboom, Eric Laenen, Anindita Maiti,  Subodh Patil, Koenraad Schalm, and Marcel Vonk for discussions. RJ also wishes to thank the organizers of the \emph{Theory $+$ AI Workshop 2025} at the Perimeter Institute for Theoretical Physics, Canada; the \emph{International Conference on Machine Learning Physics 2026} in Okinawa, Japan; and -- together with SR -- the organizers of the \emph{Physics for AI Workshop 2025} at the University of Oxford, United Kingdom; and the \emph{Machine Learning and the Renormalization Group 2024} conference in Trento, Italy, for their stimulating hospitality during various stages of this work. 

\begin{appendices}

\section{Conditional partition function}\label{sec:norm}

Our interest is in tracking a given data point through the ensemble of networks, not integrating over an ensemble of data points. Hence, the object we want to construct is the partition function \emph{conditioned on the data}, i.e., denoting the sources by $j$ (and implicitly adding such a term in the exponential),
\begin{equation}
	Z[j|x] = \int\!\mathrm{d}h_L p(h_L|x)~.
\end{equation}
Note that as a function of $x$, this is not canonically normalized in sense that $Z[0|x]\neq1$. Rather, the normalization condition is
\begin{equation}
	Z[0]=\int\!\mathrm{d}x\,Z[0|x]\,p(x)=1~,
	\label{eq:norm1}
\end{equation}
where $p(x)$ is the distribution over the data. This implies that expectation values computed with respect to $Z[0|x]$ are not c-numbers, but rather functions of $x$. 
Note that this does not affect the argument (cf. below eq. (2.11) in \cite{Grosvenor:2021eol}) that adding a source term for the auxilliary fields does not alter the partition function, since this again merely shifts the solution $y_\ell$ without altering the normalization.

In more detail, the normalization condition descends from that on the total, joint distribution:
\begin{equation}
	\begin{aligned}
		1=&\int\!\prod_{\ell=0}^L\mathrm{d}h_\ell\,p(h_L,h_{L-1},\ldots,h_1,h_0)\\
		=&\int\!\prod_{\ell=0}^L\mathrm{d}h_\ell\,p(h_L|h_{L-1},\ldots,h_0)p(h_{L-1}|h_{L-2},\ldots,h_0)\ldots p(h_2|h_1,h_0) p(h_1|h_0)p(h_0)\\
		=&\int\!\prod_{\ell=0}^L\mathrm{d}h_\ell\,p(h_L|h_{L-1})p(h_{L-1}|h_{L-2})\ldots p(h_1|h_0)p(h_0)
	\end{aligned}
\end{equation}
where the first step is the chain rule of probability, and the second step follows from the Markov property of the layers, i.e., $h_\ell$ is conditionally dependent only on $h_{\ell-1}$. We remind the reader that the data is $h_0=x$, while $h_L$ is the last hidden layer (not the output layer). Thus, the partition function -- properly normalized to unity -- is the entire right-hand side. The conditional distribution \eqref{eq:prob1} was then implicitly defined by splitting the integral such that
\begin{equation}
	\begin{aligned}
		1=&\int\!\mathrm{d}h_L\mathrm{d}x\underbrace{\int\!\prod_{\ell=1}^{L-1}\mathrm{d}h_\ell\,p(h_L|h_{L-1})p(h_{L-1}|h_{L-2})\ldots p(h_1|h_0)}_{p(h_L|x)}p(x)\\
		=&\int\!\mathrm{d}x\,p(x)\underbrace{\int\!\mathrm{d}h_L\,p(h_L|x)}_{Z[0|x]}~.
	\end{aligned}
\end{equation}
This is relevant when considering the expectation value at the initial boundary condition, $\Delta_0^i=\left<(x^i)^2\right>$, cf. below \eqref{eq:hhrec}. This expectation value is computed with respect to the ensemble defined by $Z[0|x]$; formally, if one were to exactly compute the integral over all other fields appearing in that expression, the result would be
\begin{equation}
	\left<(x^i)^2\right>=\int\!\prod_{\ell=1}^{L}\left\{\int\!\mathrm{d}h_\ell\mathrm{d}\frak{A}_\ell\int_{-i\infty}^{i\infty}\!\frac{\mathrm{d}z_\ell \mathrm{d}A_\ell}{2\pi i}\right\}\left(x^i\right)^2e^{S_0+S_\mathrm{int}}
	=(x^i)^2f(x)
\end{equation}
for some complicated function $f(x)$. In practice, $\Delta_0^i$ serves as a boundary condition that must be specified (e.g., empirically) when recursively computing the variance propagation through the network. 


\section{Analytical structure of the exact correlator}\label{sec:valiant}

In this appendix, we analyze the recursive expression \eqref{eq:Xfull} for the exact propagator, including all $O(1)$ corrections from fluctuations in the ensemble of networks, reproduced here for convenience:
\begin{equation}
	X_{\ell}=\Delta_\ell+\frac{a^2\sigma_w^2}{2}\sum_{k=1}^\infty\frac{\gamma_{2k}}{(2k)!}(2\Delta_{\ell-1})^k\sum_{n=0}^k(k-n)!\left(\frac{X_{\ell-1}}{2\Delta_{\ell-1}}\right)^n~.
	\label{eq:Xfullappx}
\end{equation}
More specifically, we first prove that this series converges for fixed arguments, and attempt to find a closed-form solution. While we have not succeeded in the latter, our investigations reveal an interesting pole structure, and we obtain a reorganization of the summation with a different interpretation when truncating to a finite number of terms, as discussed below. Thus we include this appendix in case it may be of interest to those in, for example, the resurgence or amplitudes communities, and as a report on partial progress for future work. 

\subsection*{Proof of convergence} 

We first check whether the infinite sum over $k$ converges. A common diagnostic for such factorial power series is checking the Gevrey class \cite{Marino:ResurgenceCourse}. We recall that a formal power series
\begin{equation}
	F(x)=\sum_{k=0}^\infty a_kx^k~,
	\label{eq:Gevpow}
\end{equation}
is called \emph{Gevrey-s} if there exists constants $M,b>0$ such that
\begin{equation}
	|a_k|\leq Mb^k(k!)^s
\end{equation}
for all $k\!>\!0$. With this in mind, it is convenient to first examine the asymptotic behaviour of the coefficients $\gamma_{2k}$, which were defined in \eqref{eq:gprime} based on the coefficients $g_{2k}$ given in \eqref{eq:gcoef}; since these differ only for $k=0$, we need consider only the latter, which we also reproduce here for convenience:
\begin{equation}
	g_{2k}\coloneqq a^2c_{2k}+2af_{k}~,
\end{equation}
where, for all positive integers $k$,
\begin{equation}
	f_k=\frac{2^{2k}\left(2^{2k}-1\right)B_{2k}}{(2k)!}~,
\end{equation}
and
\begin{equation}
	c_{2k}=a^22^{2k+2}\sum_{r\textrm{ odd}}^{2k}\frac{\left(2^{r+1}-1\right)\left(2^{{2k}-r+1}-1\right)B_{r+1}B_{2k-r+1}}{(r+1)!(2k-r+1)!}~.
\end{equation}
The Bernoulli numbers appearing here admit a definition in terms of the Riemann zeta function $\zeta$,
\begin{equation}
	B_{2k}=\frac{(-1)^{k+1}2(2k)!}{(2\pi)^{2k}}\zeta(2k)~,
	\qquad\quad\forall k\in\mathbb{Z}^+~.
	\label{eq:Bzeta}
\end{equation}
Given that $\zeta\!\to\!1$ as $k\!\to\!\infty$, we readily see that $f_k$ scales asymptotically like $(2/\pi)^{2k}$. Similarly, by splitting the sum over $r$ in $c_{2k}$ into two regimes ($r\ll 2k$ and $r\sim 2k$), one finds again the same scaling $(2/\pi)^{2k}$ as $k\to\infty$.\footnote{For the second regime, it is important to note that the maximum value of $r$ is $2k\!-\!1$, so that the zeta function from the second Bernoulli number has the behaviour $\zeta(2k-r+1)\to\zeta(2)=\pi^2/6$.} Thus, for fixed $a$, we have that the coefficients $\gamma_{2k}$ appearing in \eqref{eq:Xfullappx} scale like
\begin{equation}
	|\gamma_{2k}|\sim \left(\frac{4}{\pi^2}\right)^{\!k}~,
	\label{eq:gammascale}
\end{equation}
at large $k$, so that if they described a series of the form \eqref{eq:Gevpow}, it would be Gevrey-0. 

With this in hand, let us focus our attention to the inner summation in \eqref{eq:Xfullappx}, which is of the form
\begin{equation}
	\sum_{n=0}^k(k-n)!\,x^n~,
	\label{eq:innersum}
\end{equation}
where $x\coloneqq X_{\ell-1}/(2\Delta_{\ell-1})$, and examine how this quantity scales as $k\to\infty$. We do this by splitting it into three regimes: $n\ll k$, $n=\alpha k$ with $0<\alpha<1$, and $n= k$.

\subsubsection*{Regime I: $n\ll k$}

First, observe that for fixed $n\ll k$, we have
\begin{equation}
	(k-n)!=\frac{k!}{k(k-1)\cdots(k-n+1)}\sim\frac{k!}{k^n}~.
\end{equation}
as $k\to\infty$. Consequently,
\begin{equation}
	(k-n)!\,x^n\sim	k!\left(\frac{x}{k}\right)^n~.
\end{equation}
Thus, in this regime, contributions to the full sum over $k$ in \eqref{eq:Xfullappx} behave as
\begin{equation}
	\frac{\gamma_{2k}}{(2k)!}(\Delta_{\ell-1})^k(k-n)!\,x^n
	\sim\gamma_{2k}(2\Delta_{\ell-1})^k\frac{k!}{(2k)!}\left(\frac{x}{k}\right)^n~.
	\label{eq:regime1}
\end{equation}
Recalling Stirling's approximation, 
\begin{equation}
	k!\sim\sqrt{2\pi k}\left(\frac{k}{e}\right)^k~,
	\label{eq:Stirling}
\end{equation}
the ratio of factorials becomes
\begin{equation}
	\frac{k!}{(2k)!}\sim\frac{1}{\sqrt{2}}\left(\frac{e}{4k}\right)^k~,
\end{equation}
and thus, together with \eqref{eq:gammascale}, the right-hand side of \eqref{eq:regime1} scales like
\begin{equation}
	|\gamma_{2k}|(2\Delta_{\ell-1})^k\left(\frac{e}{4k}\right)^k\left(\frac{x}{k}\right)^n
	\sim \left(\frac{2\Delta_{\ell-1}e}{\pi^2k}\right)^k\frac{1}{k^n}
\end{equation}
which decays rapidly with $k$, and hence this regime is convergent.

\subsubsection*{Regime II: $0<n<k$}

In the intermediate regime, we set $n\!=\!\alpha k$ with $0\!<\!\alpha\!<\!1$. Then terms in the inner summation \eqref{eq:innersum} take the form
\begin{equation}
	(k-n)!\,x^n=[(1-\alpha)k]!\,x^{\alpha k}~,
\end{equation}
so that the contributions to the outer summation behave like
\begin{equation}
	\frac{|\gamma_{2k}|}{(2k)!}(2\Delta_{\ell-1})^k[(1-\alpha)k]!\,x^{\alpha k}
	\sim\left[\frac{2\Delta_{\ell-1}(1-\alpha)^{1-\alpha}x^\alpha}{\pi^2}\right]^k
	\left(\frac{e}{k}\right)^{(1+\alpha)k}~,
\end{equation}
via Stirling's formula \eqref{eq:Stirling} and \eqref{eq:gammascale}. This is again convergent since $1\!+\!\alpha>0$, and hence the factor of $k^{-(1+\alpha)k}$ overwhelms the growth of $\beta^k$ for any fixed $\beta$ as $k\to\infty$.

\subsubsection*{Regime III: $n= k$}

Finally, for the special case where the index $n$ reaches the maximum value of $k$, i.e., $n=k\to\infty$, we have only a single term in the inner summation \eqref{eq:innersum}, namely
\begin{equation}
	(k-k)!x^k=\left(\frac{X_{\ell-1}}{2\Delta_{\ell-1}}\right)^k~.
\end{equation}
Hence the scaling of the corresponding contributions to \eqref{eq:Xfullappx} take the simple form
\begin{equation}
	\frac{|\gamma_{2k}|}{(2k)!}X_{\ell-1}^k\sim\frac{1}{\sqrt{k}}\left(\frac{2X_{\ell-1}e}{\pi^2k}\right)^{\!k}~,
\end{equation}
whose asymptotic behaviour is again completely dominated by the decay of $k^{-k}$ as $k\to\infty$.\\

Thus, we conclude that the right-hand side of \eqref{eq:Xfullappx} converges for any fixed $\Delta_{\ell-1}$ and $X_{\ell-1}$. However, it is important to note that the recursive expression for the bare propagator in \eqref{eq:hhrec} is itself only strictly valid within the radius of convergence of the series expansion of $\mathrm{tanh}$, cf. \eqref{eq:tanhexp}. Therefore, care is needed when combining both recursive expressions, or else one must use the non-expanded form \eqref{eq:recursivehh} for the bare propagator which predicts only that in the next layer. 

Given that the series \eqref{eq:Xfullappx} is in principle tractable, we now proceed to seek a closed-form solution. As mentioned above, we have not fully succeeded in this, but our attempts reveal an interesting pole structure, and may serve as a staging area for future work.

\subsection*{Not-quite-full computation}

We begin with the inner summation in \eqref{eq:Xfullappx}, reversing the index via $m=k-n$, so that
\begin{equation}
	\begin{aligned}
		{}&\sum_{n=0}^k(k-n)!\left(\frac{X_{\ell-1}}{2\Delta_{\ell-1}}\right)^n
		=\left(\frac{X_{\ell-1}}{2\Delta_{\ell-1}}\right)^k\sum_{m=0}^k m!\left(\frac{2\Delta_{\ell-1}}{X_{\ell-1}}\right)^m\\
		  &=-\left(\frac{X_{\ell-1}}{2\Delta_{\ell-1}}\right)^{k+1}e^{-\frac{X_{\ell-1}}{2\Delta_{\ell-1}}}\left[\Gamma\left(0,-\frac{X_{\ell-1}}{2\Delta_{\ell-1}}\right)+(-1)^k(k+1)!\,\Gamma\left(-k\!-\!1,-\frac{X_{\ell-1}}{2\Delta_{\ell-1}}\right)\right]~,
	\end{aligned}
\end{equation}
where $\Gamma(s,x)$ is the (upper) incomplete Gamma function,
\begin{equation}
	\Gamma(s,x)\coloneqq\int_x^\infty\mathrm{d}t\,t^{s-1}e^{-t}~.
	\label{eq:incGamma}
\end{equation}
Explicitly, \eqref{eq:Xfullappx} is then
\begin{equation}
	X_\ell=\Delta_\ell-\frac{a^2\sigma_w^2}{2}\,Y_{\ell-1}e^{-Y_{\ell-1}}\sum_{k=1}^\infty\frac{\gamma_{2k}}{(2k)!}X_{\ell-1}^k\int_{-Y_{\ell-1}}^\infty\!\mathrm{d}t\,\frac{e^{-t}}{t}\left[1+\frac{(-1)^k(k+1)!}{t^{k+1}}\right]
	\label{eq:ohboy}
\end{equation}
where for compactness we have defined $Y_{\ell-1}\coloneqq\tfrac{X_{\ell-1}}{2\Delta_{\ell-1}}$. Taking care of the nonlinearity coefficient $\gamma_2$ as per \eqref{eq:gprime} and interchanging summation and integration order, this becomes
\begin{equation}
	\begin{aligned}
		X_\ell
			=\Delta_\ell-\frac{a^2\sigma_w^2}{2}\,Y_{\ell-1}e^{-Y_{\ell-1}}\!\int_{-Y_{\ell-1}}^\infty\!\mathrm{d}t\,\frac{e^{-t}}{t}&\Bigg\{X_{\ell-1}\left(\frac{1}{2}-\frac{1}{t^2}\right)\\
																	       &+\sum_{k=1}^\infty\frac{g_{2k}}{(2k)!}X_{\ell-1}^k\left[1+\frac{(-1)^k(k+1)!}{t^{k+1}}\right]\Bigg\}
	\end{aligned}
	\label{eq:herewego}
\end{equation}
We now focus our attention on the remaining sum over loops $k$. Substituting in \eqref{eq:gcoef} for the coefficients $g_k$, we may write this as a linear combination of four separate summations:
\begin{equation}
	\sum_{k=1}^\infty\frac{g_{2k}}{(2k)!}X_{\ell-1}^k\left[1+\frac{(-1)^k(k+1)!}{t^{k+1}}\right]
	=\sum_{k=1}^\infty\frac{X_{\ell-1}^k}{(2k)!}\left(a^2c_{2k}+2af_k\right)\left(1+\frac{(-1)^k(k+1)!}{t^{k+1}}\right)~,
	\label{eq:quadfun}
\end{equation}
where the coefficients $f$ and $c$ were defined in \eqref{eq:fcoef} and \eqref{eq:ccoef}, respectively. Since the former is simpler, we proceed with the two corresponding summations (obtained by cross-multiplying in the expression above) first. 

\subsubsection*{f-series 1}

From the first cross-term in \eqref{eq:quadfun}, we have
\begin{equation}
	\begin{aligned}
		2a\sum_{k=1}^\infty\frac{X_{\ell-1}^k}{(2k)!}f_k
		=2a\sum_{k=1}^\infty X_{\ell-1}^k\,\frac{2^{2k}\left(2^{2k}-1\right)}{(2k)!(2k)!}B_{2k}~,
	\end{aligned}
\end{equation}
To evaluate this, we note the following useful identities: first, the expression for the Bernoulli numbers in terms of the Riemann zeta function $\zeta$ given in \eqref{eq:Bzeta}, and second, that the Riemann zeta function is given by the Dirichlet series
\begin{equation}
	\zeta(s)=\sum_{n=1}^\infty\frac{1}{n^s}~,
	\qquad\quad\forall\,\mathrm{Re}(s)>1~,
	\label{eq:ZDir}
\end{equation}
which, notably, is convergent. Hence, upon writing $B_{2k}$ in the form \eqref{eq:Bzeta}, and then replacing $\zeta$ in the resulting expression by \eqref{eq:ZDir}, one obtains
\begin{equation}
	\begin{aligned}
		2a&\sum_{k=1}^\infty X_{\ell-1}^k\,\frac{2^{2k}\left(2^{2k}-1\right)}{(2k)!(2k)!}B_{2k}
		=4a\sum_{n=1}^\infty\sum_{k=1}^\infty (-1)^{k+1}\frac{2^{2k}\left(2^{2k}-1\right)}{(2k)!}\left(\frac{X_{\ell-1}}{4\pi^2 n^2}\right)^{k}\\
		   &=4a\sum_{n=1}^\infty\sum_{k=1}^\infty \frac{(-1)^{k+1}}{(2k)!}\left[\left(\frac{4X_{\ell-1}}{\pi^2 n^2}\right)^{k}-\left(\frac{X_{\ell-1}}{\pi^2 n^2}\right)^{k}\right]~,
	\end{aligned}
	\label{eq:f1temp}
\end{equation}
where we have used convergence of the Dirichlet series to interchange the summation order. Each summation over $k$ appearing in this expression is then a tractable series of the form
\begin{equation}
	\sum_{k=1}^\infty (-1)^{k+1}\frac{x^k}{(2k)!}=1-\cos\sqrt{x}~.
	\label{eq:tractor}
\end{equation}
We thus obtain
\begin{equation}
	\begin{aligned}
		2a\sum_{k=1}^\infty\frac{X_{\ell-1}^k}{(2k)!}f_k
		=4a\sum_{n=1}^\infty \left(\cos\sqrt{y_n}-\cos\sqrt{4y_n}\right)
	\end{aligned}
	\label{eq:fseries1}
\end{equation}
where for present and future compactness we have defined
\begin{equation}
	y_n\coloneqq \frac{X_{\ell-1}}{\pi^2 n^2}~.
	\label{eq:YX}
\end{equation}

\subsubsection*{f-series 2}

The second summation involving $f$ coefficients is then very similar to the first, except that it picks up the last factor in \eqref{eq:quadfun}; hence we can modify \eqref{eq:f1temp} to obtain
\begin{equation}
	\begin{aligned}
		\frac{2a}{t}\sum_{k=1}^\infty&\frac{(-1)^k(k+1)!}{(2k)!}\left(\frac{X_{\ell-1}}{t}\right)^k f_k
		=-\frac{4a}{t}\sum_{n=1}^\infty\sum_{k=1}^\infty \frac{(k+1)!}{(2k)!}\left[\left(\frac{4y_n}{t}\right)^{k}-\left(\frac{y_n}{t}\right)^{k}\right]\\
		=-\frac{4a}{t}&\sum_{n=1}^\infty\Bigg[
			\frac{3y_n}{4t}
			-\left(\frac{3}{2}+\frac{y_n}{4t}\right)\sqrt{\frac{\pi y_n}{4t}}\,e^{\frac{y_n}{4t}}\,\mathrm{erf}\sqrt{\frac{y_n}{4t}}
			+\left(\frac{3}{2}+\frac{y_n}{t}\right)\sqrt{\frac{\pi y_n}{t}}\,e^{\frac{y_n}{t}}\,\mathrm{erf}\sqrt{\frac{y_n}{t}}
			\,\Bigg]
	\end{aligned}
	\label{eq:fseries2}
\end{equation}
with $y_n$ defined in \eqref{eq:YX}, and where 
\begin{equation}
	\mathrm{erf}(z)\coloneqq\frac{2}{\sqrt{\pi}}\int_0^z\!\mathrm{d}\rho\,e^{-\rho^2}
	\label{eq:erf}
\end{equation}
is the error function.

\subsubsection*{c-series 1}

Now for the two summations involving $c$ coefficients \eqref{eq:ccoef}. The first is
\begin{equation}
	\begin{aligned}
		a^2&\sum_{k=1}^\infty\frac{X_{\ell-1}^k}{(2k)!}c_{2k}
		=4a^2\sum_{k=1}^\infty\frac{(4X_{\ell-1})^k}{(2k)!}\sum_{r\textrm{ odd}}^{2 k}\frac{\left(2^{r+1}-1\right)\left(2^{2k-r+1}-1\right)}{(r+1)!(2k-r+1)!}B_{r+1}B_{2k-r+1}\\
		    &=\frac{4a^2}{\pi^2}\sum_{k=1}^\infty\frac{(-1)^{k+1}}{(2k)!}\left(\frac{X_{\ell-1}}{\pi^2}\right)^{\!k}\sum_{r \textrm{ odd}}^{2 k}\left(2^{r+1}-1\right)\left(2^{2k-r+1}-1\right)\zeta(r+1)\zeta(2k-r+1)\\
		   &=\frac{4a^2}{\pi^2}\sum_{m,n=1}^\infty\frac{1}{mn}\sum_{k=1}^\infty\frac{(-1)^{k+1}}{(2k)!}\left(\frac{X_{\ell-1}}{\pi^2 n^2}\right)^{\!k}\sum_{r\textrm{ odd}}^{2k}\left(\frac{n}{m}\right)^r\left(2^{r+1}-1\right)\left(2^{2k-r+1}-1\right)
	\end{aligned}
	\label{eq:c1hell}
\end{equation}
where in the first step, we applied \eqref{eq:Bzeta} to the product of Bernoulli numbers, and in the second we expressed the zeta functions in terms of their Dirichlet series \eqref{eq:ZDir} and used the convergence thereof to swap the summation orders. Expanding the factors within the innermost summation then leads to a linear combination of simple geometric series:
\begin{equation}
	\begin{aligned}
		\sum_{r\,\mathrm{odd}}^{2k}&\left(\frac{n}{m}\right)^r\left(2^{r+1}-1\right)\left(2^{2k-r+1}-1\right)\\
					      &=\left(2^{2k+2}+1\right)\sum_{r\,\mathrm{odd}}^{2k}\left(\frac{n}{m}\right)^r
					      -2^{2k+1}\sum_{r\,\mathrm{odd}}^{2k}\left(\frac{n}{2m}\right)^r
					      -2\sum_{r\,\mathrm{odd}}^{2k}\left(\frac{2n}{m}\right)^r~.
	\end{aligned}
	\label{eq:geofun}
\end{equation}
Each of these requires a case-by-case analysis depending on the value of the ratio:
\begin{equation}
	\begin{aligned}
		\left(2^{2k+2}+1\right)\sum_{r\,\mathrm{odd}}^{2k}\left(\frac{n}{m}\right)^r&=\left(2^{2k+2}+1\right)\times
		\begin{cases}
			\frac{mn}{n^2-m^2}\left(\left(\frac{n}{m}\right)^{2k}-1\right)~,\qquad &m\neq n~,\\
			k~,\qquad& m=n~,
		\end{cases}\\
		-2^{2k+1}\sum_{r\,\mathrm{odd}}^{2k}\left(\frac{n}{2m}\right)^r&=-2^{2k+1}\times
		\begin{cases}
			\frac{2mn}{n^2-4m^2}\left(\left(\frac{n}{2m}\right)^{2k}-1\right)~,\qquad&2m\neq n~,\\
			k\qquad& 2m=n~,
		\end{cases}\\
		-2\sum_{r\,\mathrm{odd}}^{2k}\left(\frac{2n}{m}\right)^r&=
		\begin{cases}
			\frac{-4mn}{4n^2-m^2}\left(\left(\frac{2n}{m}\right)^{2k}-1\right)~,\qquad& m\neq2n~,\\
			-2k~,\qquad& m=2n~.
		\end{cases}
	\end{aligned}
\end{equation}\label{C27}
In the parent expression \eqref{eq:c1hell} therefore, we organize the outermost sums over $n$ and $m$ such that, schematically,
\begin{equation}
	\frac{4a^2}{\pi^2}\sum_{n,m=1}^\infty(\ldots)
	=\frac{4a^2}{\pi^2}\left[\sum_{n=m}+\sum_{n=2m}+\sum_{2n=m}+\sum_{n\neq Cm}\right]~,
\end{equation}
where $C\in\{1/2,1,2\}$ (and we suppress the ellipses on the right-hand side for compactness). Then in the first three terms, the internal sums over $k$ reduce to sums of the form \eqref{eq:tractor}, which we can subsequently collect to obtain
\begin{equation}
	\begin{aligned}
		{}&\left[\sum_{n=m}+\sum_{n=2m}+\sum_{2n=m}\right]\frac{1}{mn}\sum_{k=1}^\infty\frac{(-1)^{k+1}}{(2k)!}y_n^k\sum_{r\textrm{ odd}}^{2k}\left(\frac{n}{m}\right)^r\left(2^{r+1}-1\right)\left(2^{2k-r+1}-1\right)\\
		  &\qquad=\sum_{n=1}^\infty\frac{1}{15n^2}\left[\cos\sqrt{\frac{y_n}{4}}-21\cos\sqrt{y_n}+84\cos\sqrt{4y_n}-64\cos\sqrt{16y_n}\right]~.
	\end{aligned}
	\label{eq:firstthree}
\end{equation}
Meanwhile, for the final term, we obtain
\begin{equation}
	\begin{aligned}
		\sum_{n\neq Cm}&\frac{1}{mn}\sum_{k=1}^\infty\frac{(-1)^{k+1}}{(2k)!}y_n^k\sum_{r\textrm{ odd}}^{2k}\left(\frac{n}{m}\right)^r\left(2^{r+1}-1\right)\left(2^{2k-r+1}-1\right)\\
		  &=\sum_{n\neq Cm}\frac{2}{m^2-n^2}\left[\frac{3m^2\cos\sqrt{y_n}}{m^2-4n^2}-\frac{12m^2\cos\sqrt{4y_n}}{4m^2-n^2}\right]\\
		  &=2\!\sum_{n\neq Cm}\left[\frac{-\cos\sqrt{y_n}-4\cos\sqrt{4y_n}}{m^2-n^2}+\frac{4\cos\sqrt{y_n}}{m^2-4n^2}+\frac{4\cos\sqrt{4y_n}}{4m^2-n^2}\right]~,
	\end{aligned}
	\label{eq:nCm1}
\end{equation}
where in the second line, we used the freedom to interchange dummy indices $m\longleftrightarrow n$ to combine terms, and in the third we have decomposed the result into partial fractions for what follows. The remaining sums over $n,m$ explicitly exclude the poles at $m=n/2$, $n$, and $2n$. Thus, we can evaluate the sum over $m$ by using the well-known partial fraction expansion for $\cot$,
\begin{equation}
    \sum_{m=1}^{\infty}\frac{1}{m^2-z^2}=\frac{1}{2z^2}-\frac{\pi}{2z}\cot(\pi z)~,
	\label{eq:cotangent}
\end{equation}
and carefully subtracting the pole as $z\to n\in\mathbb{Z}$. Explicitly, consider the first term in \eqref{eq:nCm1}, with a pole at $m=n$. The pole-subtracted sum over $m$ is given by
\begin{equation}
	\begin{aligned}
		-2\left(\cos\sqrt{y_n}+4\cos\sqrt{4y_n}\right)&\lim_{z\to n}\left[\sum_{m=1}^\infty\frac{1}{m^2-z^2}-\frac{1}{n^2-z^2}\right]\\
		=-2\left(\cos\sqrt{y_n}+4\cos\sqrt{4y_n}\right)&\lim_{z\to n}\left[\frac{1}{2z^2}-\frac{\pi}{2 z}\cot(\pi z)-\frac{1}{n^2-z^2}\right]~.
	\end{aligned}
\end{equation}
To perform the subtraction in a controlled manner, we take $z=n+\epsilon$ and expand for small $\epsilon$:
\begin{equation}
	\begin{aligned}
		{}&\lim_{\epsilon\to 0}\left[\frac{1}{2(n+\epsilon)^2}-\frac{\pi}{2 (n+\epsilon)}\cot\left(\pi(n+\epsilon)\right)-\frac{1}{n^2-(n+\epsilon)^2}\right]
		=\frac{3}{4n^2}
	\end{aligned}
\end{equation}
where we used the periodicity of $\cot$ for integer $n$, and expanded ${\cot(\pi \epsilon)=(\pi \epsilon)^{-1}-\pi \epsilon/3+O(\epsilon^2)}$. Similarly, for the second term in \eqref{eq:nCm1}, we subtract the pole at $m=2n$ via
\begin{equation}
	8\cos\sqrt{y_n}\lim_{z\to n}\left[\sum_{m=1}^\infty\frac{1}{m^2-4z^2}-\frac{1}{4n^2-4z^2}\right]=\frac{3}{2 n^2}\cos\sqrt{y_n}~.
\end{equation}
Finally, the third term in \eqref{eq:nCm1} is slightly tricky, since the periodicity relation ${\cot\left(\tfrac{\pi}{2}(n+\epsilon)\right)=\cot\left(\tfrac{\pi}{2}\epsilon\right)}$ holds only for even values of $n$; for odd values, the series has no pole since $m=n/2$ is not an integer. Thus, we implicitly split the outer sum over $n$ into even and odd integers, and perform the pole subtraction only for the former:
\begin{equation}
	n\textrm{ even:}\qquad 2\cos\sqrt{4y_n}\lim_{z\to n}\left[\sum_{m=1}^\infty\frac{1}{m^2-z^2/4}-\frac{4}{n^2-z^2}\right]=\frac{6}{n^2}\cos\sqrt{4y_n}~.
\end{equation}
Meanwhile, for $n$ odd, the $\cot$ in \eqref{eq:cotangent} vanishes, and we obtain
\begin{equation}
	n\textrm{ odd:}\qquad 2\cos\sqrt{4y_n}\sum_{m=1}^\infty\frac{1}{m^2-n^2/4}=\frac{2}{n^2}-\frac{\pi}{n}\cancel{\cot\left(\frac{\pi n}{2}\right)}=\frac{4}{n^2}\cos\sqrt{4y_n}~.
\end{equation}
Combining these results in \eqref{eq:nCm1}, we thus obtain
\begin{equation}
	\begin{aligned}
		\sum_{n\neq Cm}&\frac{1}{mn}\sum_{k=1}^\infty\frac{(-1)^{k+1}}{(2k)!}y_n^k\sum_{r\textrm{ odd}}^{2k}\left(\frac{n}{m}\right)^r\left(2^{r+1}-1\right)\left(2^{2k-r+1}-1\right)\\
		=&\sum_{n=1}^\infty\left[-\frac{3}{2n^2}\left(\cancel{\cos\sqrt{y_n}}+4\cos\sqrt{4y_n}\right)+\frac{3}{2n^2}\cancel{\cos\sqrt{4y_n}}\right]\\
		 &+6\sum_{n\,\textrm{even}}\frac{\cos\sqrt{4y_n}}{n^2}+4\sum_{n\,\textrm{odd}}\frac{\cos\sqrt{4y_n}}{n^2}\\
		=&-6\sum_{n=1}^\infty\frac{\cos\sqrt{4y_n}}{n^2}+6\sum_{n\,\textrm{even}}\frac{\cos\sqrt{4y_n}}{n^2}+4\sum_{n\,\textrm{odd}}\frac{\cos\sqrt{4y_n}}{n^2}\\
		=&-2\sum_{n\,\textrm{odd}}\frac{\cos\sqrt{4y_n}}{n^2}~,
	\end{aligned}
\end{equation}
where in these expressions it is understood that $n\geq1$. Together with the first three terms \eqref{eq:firstthree}, we thus obtain
\begin{equation}
	\begin{aligned}
	a^2\sum_{k=1}^\infty\frac{X_{\ell-1}^k}{(2k)!}c_{2k}
	=&\frac{4a^2}{15\pi^2}\sum_{n=1}^\infty\frac{1}{n^2}\left[\cos\sqrt{\frac{y_n}{4}}-21\cos\sqrt{y_n}+84\cos\sqrt{4y_n}-64\cos\sqrt{16y_n}\right]\\
	&-\frac{8a^2}{\pi^2}\sum_{n\,\textrm{odd}}\frac{1}{n^2}\cos\sqrt{4y_n}~.
	\end{aligned}
	\label{eq:cseries1}
\end{equation}

\subsubsection*{c-series 2}

Finally, the second summation involving $c$ coefficients is obtained from the above by including the last factor in \eqref{eq:quadfun}. Since this affects only the sums over $k$, we immediately have
\begin{equation}
	\begin{aligned}
		{}&\frac{a^2}{t}\sum_{k=1}^\infty\frac{(-1)^k(k+1)!}{(2k)!}\left(\frac{X_{\ell-1}}{t}\right)^kc_{2k}\\
		  &=-\frac{4a^2}{t\pi^2}\sum_{m,n=1}^\infty\frac{1}{mn}\sum_{k=1}^\infty\frac{(k+1)!}{(2k)!}\left(\frac{y_n}{t}\right)^k\sum_{r\textrm{ odd}}^{2k}\left(\frac{n}{m}\right)^r\left(2^{r+1}-1\right)\left(2^{2k-r+1}-1\right)~,
	\end{aligned}
	\label{eq:c2hell}
\end{equation}
cf. \eqref{eq:c1hell}. Performing the same case-by-case analysis for the geometric series (over $r$) as above, and then evaluating the sums over $k$, we find
\begin{equation}
	\begin{aligned}
		{}\Bigg[\sum_{n=m}+\sum_{n=2m}&+\sum_{2n=m}\Bigg]\frac{1}{mn}\sum_{k=1}^\infty\frac{(k+1)!}{(2k)!}\left(\frac{y_n}{t}\right)^k\sum_{r\textrm{ odd}}^{2k}\left(\frac{n}{m}\right)^r\left(2^{r+1}-1\right)\left(2^{2k-r+1}-1\right)\\
		=\sum_{n=1}^\infty\frac{1}{15n^2}&\bigg[\frac{2835y_n}{16t}
			-\left(\frac{3}{2}+\frac{y_n}{16 t}\right)\sqrt{\frac{\pi y_n}{ 16 t}}\,e^{\frac{y_n}{16 t}}\,\mathrm{erf}\sqrt{\frac{y_n}{16t}}
			+21\left(\frac{3}{2}+\frac{y_n}{4 t}\right)\sqrt{\frac{\pi y_n}{4 t}}\,e^{\frac{y_n}{4 t}}\,\mathrm{erf}\sqrt{\frac{y_n}{4t}}\\
			&-84\left(\frac{3}{2}+\frac{y_n}{t}\right)\sqrt{\frac{\pi y_n}{t}}\,e^{\frac{y_n}{t}}\,\mathrm{erf}\sqrt{\frac{y_n}{t}}
			+64\left(\frac{3}{2}+\frac{4y_n}{t}\right)\sqrt{\frac{4\pi y_n}{t}}\,e^{\frac{4y_n}{t}}\,\mathrm{erf}\sqrt{\frac{4y_n}{t}}\bigg]~,
	\end{aligned}
\end{equation}
similar to \eqref{eq:fseries2}, where the error function is given in \eqref{eq:erf}. Meanwhile, for the $n\neq Cm$ terms, we have
\begin{equation}
	\begin{aligned}
		{}\sum_{n\neq Cm}\frac{1}{mn}&\sum_{k=1}^\infty\frac{(k+1)!}{(2k)!}\left(\frac{y_n}{t}\right)^k\sum_{r\textrm{ odd}}^{2k}\left(\frac{n}{m}\right)^r\left(2^{r+1}-1\right)\left(2^{2k-r+1}-1\right)\\
		=\sum_{n\neq Cm}&\bigg[\frac{1}{m^2}\frac{9y_n}{4t}
					 -\frac{6m^2}{m^4-5m^2n^2+4n^4}\left(\frac{3}{2}+\frac{y_n}{4 t}\right)\sqrt{\frac{\pi y_n}{4 t}}\,e^{\frac{y_n}{4 t}}\,\mathrm{erf}\sqrt{\frac{y_n}{4 t}}\\
				 &+\frac{24m^2}{4m^4-5m^2n^2+n^4}\left(\frac{3}{2}+\frac{y_n}{t}\right)\sqrt{\frac{\pi y_n}{t}}\,e^{\frac{y_n}{t}}\,\mathrm{erf}\sqrt{\frac{y_n}{t}}
				 \bigg]\\
		=\sum_{n\neq Cm}&\bigg\{\frac{1}{m^2}\frac{9y_n}{4t}
			+\frac{1}{m^2-n^2}\left[2\left(\frac{3}{2}+\frac{y_n}{4t}\right)\sqrt{\frac{\pi y_n}{4t}}\,e^{\frac{y_n}{4t}}\mathrm{erf}\sqrt{\frac{y_n}{4 t}}
			+8\left(\frac{3}{2}+\frac{y_n}{t}\right)\sqrt{\frac{\pi y_n}{t}}\,e^{\frac{y_n}{t}}\mathrm{erf}\sqrt{\frac{y_n}{t}}\right]\\
				&-\frac{8}{m^2-4n^2}\left(\frac{3}{2}+\frac{y_n}{4 t}\right)\sqrt{\frac{\pi y_n}{4 t}}\,e^{\frac{y_n}{4 t}}\mathrm{erf}\sqrt{\frac{y_n}{4 t}}
				-\frac{8}{4m^2-n^2}\left(\frac{3}{2}+\frac{y_n}{t}\right)\sqrt{\frac{\pi y_n}{t}}\,e^{\frac{y_n}{t}}\mathrm{erf}\sqrt{\frac{y_n}{t}}
		\bigg\}\\
	\end{aligned}
\end{equation}
where again we have used the freedom to interchange dummy indices in the second equality, and in the third we have decomposed the result into distinct poles. The latter are of precisely the same form as in c-series 1 above, cf. \eqref{eq:nCm1}, so we can re-use the pole-subtracted results for the summations over $m$ to immediately obtain
\begin{equation}
	\begin{aligned}
		\sum_{n\neq Cm}&\frac{1}{mn}\sum_{k=1}^\infty\frac{(k+1)!}{(2k)!}\left(\frac{y_n}{t}\right)^k\sum_{r\textrm{ odd}}^{2k}\left(\frac{n}{m}\right)^r\left(2^{r+1}-1\right)\left(2^{2k-r+1}-1\right)\\
			       &=\sum_{n=1}^\infty\left[\frac{3\pi^2 y_n}{8 t}+\frac{6}{n^2}\left(\frac{3}{2}+\frac{y_n}{t}\right)\sqrt{\frac{\pi y_n}{t}}\,e^{\frac{y_n}{t}}\mathrm{erf}\sqrt{\frac{y_n}{t}}\right]\\
			       &-6\sum_{n\,\mathrm{even}}\frac{1}{n^2}\left(\frac{3}{2}+\frac{y_n}{t}\right)\sqrt{\frac{\pi y_n}{t}}\,e^{\frac{y_n}{t}}\mathrm{erf}\sqrt{\frac{y_n}{t}}
			      -4\sum_{n\,\mathrm{odd}}\frac{1}{n^2}\left(\frac{3}{2}+\frac{y_n}{t}\right)\sqrt{\frac{\pi y_n}{t}}\,e^{\frac{y_n}{t}}\mathrm{erf}\sqrt{\frac{y_n}{t}}\\
			       &=\sum_{n=1}^\infty\frac{3\pi^2 y_n}{8 t}+2\sum_{n\,\mathrm{odd}}\frac{1}{n^2}\left(\frac{3}{2}+\frac{y_n}{t}\right)\sqrt{\frac{\pi y_n}{t}}\,e^{\frac{y_n}{t}}\mathrm{erf}\sqrt{\frac{y_n}{t}}~.
	\end{aligned}
\end{equation}
Thus, combining results, we obtain
\begin{equation}
	\begin{aligned}
		\frac{a^2}{t}\sum_{k=1}^\infty\frac{(-1)^k(k+1)!}{(2k)!}&\left(\frac{X_{\ell-1}}{t}\right)^kc_{2k}\\
		=-\frac{4a^2}{t\pi^2}\sum_{n=1}^\infty\frac{1}{n^2}&\bigg[\left(\frac{63}{2}+\pi^2 n^2\right)\frac{3y_n}{8t}
			-\frac{1}{15}\left(\frac{3}{2}+\frac{y_n}{16 t}\right)\sqrt{\frac{\pi y_n}{ 16 t}}\,e^{\frac{y_n}{16 t}}\,\mathrm{erf}\sqrt{\frac{y_n}{16t}}\\
								   &+\frac{7}{5}\left(\frac{3}{2}+\frac{y_n}{4 t}\right)\sqrt{\frac{\pi y_n}{4 t}}\,e^{\frac{y_n}{4 t}}\,\mathrm{erf}\sqrt{\frac{y_n}{4t}}
			-\frac{28}{5}\left(\frac{3}{2}+\frac{y_n}{t}\right)\sqrt{\frac{\pi y_n}{t}}\,e^{\frac{y_n}{t}}\,\mathrm{erf}\sqrt{\frac{y_n}{t}}\\
								   &+\frac{64}{15}\left(\frac{3}{2}+\frac{4y_n}{t}\right)\sqrt{\frac{4\pi y_n}{t}}\,e^{\frac{4y_n}{t}}\,\mathrm{erf}\sqrt{\frac{4y_n}{t}}\bigg]\\
								   &-\frac{8a^2}{t\pi^2}\sum_{n\,\mathrm{odd}}\left(\frac{3}{2}+\frac{y_n}{t}\right)\sqrt{\frac{\pi y_n}{t}}\,e^{\frac{y_n}{t}}\mathrm{erf}\sqrt{\frac{y_n}{t}}~.
	\end{aligned}
	\label{eq:cseries2}
\end{equation}

\subsubsection*{Total}

The partial results above may be naturally collected into two classes, depending on whether they contain $\cos$ or $\mathrm{erf}$: that is, \eqref{eq:fseries1} and \eqref{eq:cseries1} combined yield
\begin{equation}
	\begin{aligned}
		f1+c1=\sum_{n=1}^\infty&\bigg[\frac{4a^2}{15\pi^2n^2}\cos\sqrt{\frac{y_n}{4}}+4a\left(1-\frac{7a}{5\pi^2n^2}\right)\cos\sqrt{y_n}\\
				       &-4a\left(1-\frac{28a}{5\pi^2n^2}\right)\cos\sqrt{4y_n}-\frac{256a^2}{15\pi^2n^2}\cos\sqrt{16y_n}\bigg]\\
			-\frac{8a^2}{\pi^2}&\sum_{n\,\textrm{odd}}\frac{1}{n^2}\cos\sqrt{4y_n}~.
	\end{aligned}
	\label{eq:f1c1}
\end{equation}
while \eqref{eq:fseries2} and \eqref{eq:cseries2} combined yield
\begin{equation}
	\begin{aligned}
		f2+c2=\sum_{n=1}^\infty
	&\bigg[-\frac{12ay_n}{4t^2}\left[1+\frac{a}{2}\left(1+\frac{63}{2\pi^2n^2}\right)\right]\\
		&+\frac{4a^2}{15t\pi^2n^2}\left(\frac{3}{2}+\frac{y_n}{16 t}\right)\sqrt{\frac{\pi y_n}{ 16 t}}\,e^{\frac{y_n}{16 t}}\,\mathrm{erf}\sqrt{\frac{y_n}{16t}}\\
		      &+\frac{4a}{t}\left(1-\frac{7a}{5\pi^2n^2}\right)\left(\frac{3}{2}+\frac{y_n}{4 t}\right)\sqrt{\frac{\pi y_n}{4 t}}\,e^{\frac{y_n}{4 t}}\,\mathrm{erf}\sqrt{\frac{y_n}{4t}}\\
		&-\frac{4a}{t}\left(1-\frac{28a}{5\pi^2n^2}\right)\left(\frac{3}{2}+\frac{y_n}{t}\right)\sqrt{\frac{\pi y_n}{t}}\,e^{\frac{y_n}{t}}\,\mathrm{erf}\sqrt{\frac{y_n}{t}}\\
		&-\frac{256a^2}{15t\pi^2n^2}\left(\frac{3}{2}+\frac{4y_n}{t}\right)\sqrt{\frac{4\pi y_n}{t}}\,e^{\frac{4y_n}{t}}\,\mathrm{erf}\sqrt{\frac{4y_n}{t}}\bigg]\\
		-\frac{8a^2}{t\pi^2}&\sum_{n\,\mathrm{odd}}\left(\frac{3}{2}+\frac{y_n}{t}\right)\sqrt{\frac{\pi y_n}{t}}\,e^{\frac{y_n}{t}}\mathrm{erf}\sqrt{\frac{y_n}{t}}~.
	\end{aligned}
	\label{eq:f2c2}
\end{equation}
The total sum over loops \eqref{eq:quadfun} is thus
\begin{equation}
	\sum_{k=1}^\infty\frac{g_{2k}}{(2k)!}X_{\ell-1}^k\left[1+\frac{(-1)^k(k+1)!}{t^{k+1}}\right]
	=(f1+c1)+(f2+c2)
	\eqqcolon\blacklozenge
	\label{eq:finaldiamond}
\end{equation}
which we have denoted $\blacklozenge$ for compactness, so that we may concisely express \eqref{eq:herewego} -- the recursive expression for the exact propagator, including all loop corrections -- as
\begin{equation}
	\begin{aligned}
		X_\ell
		=\Delta_\ell-\frac{a^2\sigma_w^2}{2}\,Y_{\ell-1}e^{-Y_{\ell-1}}\!\int_{-Y_{\ell-1}}^\infty\!\mathrm{d}t\,\frac{e^{-t}}{t}&\Bigg\{X_{\ell-1}\left(\frac{1}{2}-\frac{1}{t^2}\right)+\blacklozenge\Bigg\}~,
	\end{aligned}
	\label{eq:final}
 \end{equation}
where we remind the reader that
\begin{equation}
	y_n\coloneqq \frac{X_{\ell-1}}{\pi^2 n^2}
	\qquad\mathrm{and}\qquad
	Y_{\ell-1}\coloneqq\frac{X_{\ell-1}}{2\Delta_{\ell-1}}
\end{equation}
cf. \eqref{eq:YX} and just below \eqref{eq:ohboy}.

Unfortunately, we have not managed to find a closed-form expression for either of \eqref{eq:f1c1} or \eqref{eq:f2c2}, with the exception of the first term in the latter, with no error function, which can be summed exactly:
\begin{equation}
	-\sum_{n=1}^\infty\frac{12ay_n}{4t^2}\left[1+\frac{a}{2}\left(1+\frac{63}{2\pi^2n^2}\right)\right]=-\frac{a}{t^2}\left(\frac{1}{2}+\frac{31\,a}{40}\right)X_{\ell-1}~.
\end{equation}
However, we note that the argument of the remaining, total sum over $n$ decays rapidly, and thus $\blacklozenge$ is amenable to numerical computation to high accuracy. Physically, we have traded the sum over loops and recursion nodes for a sum over the number of terms retained in the approximation to each vertex, which may be interesting because truncating to finite $k$ is not a consistent loop truncation.

That is, observe that the original sum over $k$ in \eqref{eq:quadfun} is not a na\"ive sum over loops, but a sum over the number of petals appearing in the base of the diagram, cf. \eqref{eq:flowers}. Even the first term, $k\!=\!1$, may include up to $L$ loops (where $\ell\in[0,L]$) via the recursion node that allows the repeated insertion of $k\!=\!1$ flower diagrams corresponding to sums over neurons at previous layers---see the third diagram on the right-hand side of \eqref{eq:fun}. That is, the first diagram in each of \eqref{eq:cac1}, \eqref{eq:n2}, and \eqref{eq:n3} would all be included at $k\!=\!1$ in the computation of $X_3$. For $k\!=\!2$, the situation is already much more complicated, since this will include a single tower with up to $L$ loops, as well as all possible dual towers with between 1 and $L$ loops each. Thus, it is unclear if truncating this sum to any fixed $k$ has a sensible physical interpretation. 

Conversely, the sum over $n$ that remains in \eqref{eq:finaldiamond} ultimately stems from the Dirichlet series expression for the zeta function \eqref{eq:ZDir}, which in turn arose from the Bernoulli numbers appearing in the coefficients of the Taylor expansion of the activation function $\tanh$. Truncating this to finite $n$ is therefore ultimately an approximation of these coefficients. Thus, our resummation effectively reorganizes the expression \eqref{eq:quadfun} into one in which the sum over infinitely-many loop diagrams has been performed, but which leaves a sum representing the number of terms retained in the approximation of each vertex; meanwhile the remaining integral over $t$ in \eqref{eq:final} stems from the sum over recursion nodes. This last is somewhat reminiscent of the Mellin transform that appears in the study of CFT correlators and amplitudes in AdS \cite{Fitzpatrick:2011ia,Penedones_2011,Sleight_2021,Goncalves:2014rfa}. More generally, the decomposition of propagators into spectral functions -- in particular the Dirichlet series for the zeta function that appears here -- via the heat kernel expansion is a rich subject \cite{Vassilevich_2003}. Lastly, while the convergence of the initial series makes the Borel transform itself unnecessary, there is a large body of work on extracting non-perturbative physics from the pole structure inherent in the perturbative expansion, for example in resurgence \cite{Dorigoni_2019,Marino:ResurgenceCourse}. It could be interesting to more thoroughly investigate whether the analytical structure of the exact two-point function explored here is amenable to study via these methods, but we leave this for future work.

\end{appendices}

\bibliographystyle{ytphys}
\bibliography{biblio}

\end{document}